\documentclass[10pt, journal]{IEEEtran}

\usepackage{cite}
\usepackage{amsmath,amssymb,amsfonts}
\usepackage{mathtools}
\usepackage{bm}
\usepackage{dsfont}
\DeclareMathOperator*{\E}{\mathbb{E}}
\DeclareMathOperator*{\V}{\mathbb{V}}
\usepackage{graphicx}
\usepackage{textcomp}
\usepackage{subfig}
\usepackage[font=footnotesize]{caption}
\usepackage[table]{xcolor}
\usepackage{enumitem}
\usepackage{balance}
\usepackage{multirow}
\usepackage{array}
\usepackage{adjustbox}

\usepackage{makecell}
\usepackage[linesnumbered, ruled, lined,hangingcomment]{algorithm2e}
\usepackage{siunitx}
\usepackage{pgfplots}
\pgfplotsset{compat=1.18}
\usepgfplotslibrary{groupplots}
\usepgfplotslibrary{colorbrewer}

\begin{document}
\bstctlcite{IEEEexample:BSTcontrol}

\title{Variability-Aware Approximate Circuit Synthesis via Genetic Optimization}

\author{
Konstantinos~Balaskas,
Florian~Klemme,~\IEEEmembership{Member,~IEEE,}
Georgios~Zervakis,\\
Kostas~Siozios, \IEEEmembership{Senior Member, IEEE},
Hussam~Amrouch, \IEEEmembership{Member, IEEE},
J\"org~Henkel, \IEEEmembership{Fellow, IEEE}

\thanks{Manuscript received March 09, 2022; revised June 06, 2022; accepted June 13, 2022. (\textit{Corresponding author: K. Balaskas, e-mail:konstantinos.balaskas@kit.edu}).}%
\thanks{This work is supported in parts by the German Research Foundation (DFG) project ``ACCROSS''  HE 2343/16-1, AM 534/3-1 under the grant 428566201, in part by the E.C. Funded Program ``SERRANO'' under H2020 Grant 101017168, and in part by Advantest as part of the Graduate School ``Intelligent Methods for Test and Reliability'' (GS-IMTR) at the University of Stuttgart.
}
\thanks{K. Balaskas and K. Siozios are with the Department of Physics, Aristotle University of Thessaloniki, Thessaloniki 54124, Greece. K. Balaskas is also with the Chair for Embedded Systems at Karlsruhe Institute of Technology, Karlsruhe 76131, Germany.}%
\thanks{F. Klemme and H.~Amrouch are with the Chair for Semiconductor Test and Reliability (STAR) at University of Stuttgart, Stuttgart 70174, Germany.}%
\thanks{G. Zervakis and J. Henkel are with the Chair for Embedded Systems at Karlsruhe Institute of Technology, Karlsruhe 76131, Germany.}%
}

\markboth{Published in IEEE Transactions on Circuits and Systems I: Regular Papers. DOI: 10.1109/TCSI.2022.3183858}%
{K. Balaskas \MakeLowercase{\textit{et al.}}: Variability-Aware Approximate Circuit Synthesis via Genetic Optimization}

\maketitle
\global\csname @topnum\endcsname 0
\global\csname @botnum\endcsname 0

\begin{abstract}
One of the major barriers that CMOS devices face at nanometer scale is increasing parameter variation due to manufacturing imperfections.
Process variations severely inhibit the reliable operation of circuits, as the operational frequency at the nominal process corner is insufficient to suppress timing violations across the entire variability spectrum.
To avoid variability-induced timing errors, previous efforts impose pessimistic and performance-degrading timing guardbands atop the operating frequency.
In this work, we employ approximate computing principles and propose a circuit-agnostic automated framework for generating variability-aware approximate circuits that eliminate process-induced timing guardbands.
Variability effects are accurately portrayed with the creation of variation-aware standard cell libraries, fully compatible with standard EDA tools.
The underlying transistors are fully calibrated against industrial measurements from Intel 14nm FinFET in which both electrical characteristics of transistors and variability effects are accurately captured.
In this work, we explore the design space of approximate variability-aware designs to automatically generate circuits of reduced variability and increased performance without the need for timing guardbands.
Experimental results show that by introducing negligible functional error of merely $\bm{5.3 \times 10^{-3}}$, our variability-aware approximate circuits can be reliably operated under process variations without sacrificing the application performance.
\end{abstract}


\section{Introduction}\label{sec:intro}
\begin{figure}[t]
    \centering
    \resizebox{1\columnwidth}{!}{
        \includegraphics{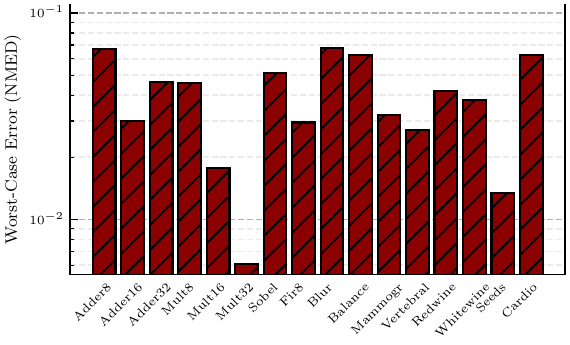}
    }
    \caption{Worst-case error evaluation obtained from a 1000-point Monte-Carlo variability analysis for several arithmetic circuits, image processing benchmarks, and machine learning classifiers (see Section~\ref{sec:evaluation}). The Normalized Mean Error Distance (NMED) error metric is used (see Section~\ref{sec:csim}).}
    \label{fig:wc_nmed}
\end{figure}

Technology scaling has brought significant challenges to circuit designers due to the intrinsic physical limitations of silicon devices~\cite{Bowman:IEEE-SSC:2010:45}.
One of the major barriers that CMOS devices face at nanometer scale is increasing process parameter variations.
Process imperfections, such as metal work-function, random dopant fluctuations and line-edge roughness cause devices to exhibit large variability in their electrical parameters, particularly in the threshold voltage $V_{th}$, as well as in other device parameters (e.g., channel length, gate width, oxide thickness)~\cite{Bhunia:VLSID:2007} and circuit wiring.
Thus, the operating frequency becomes unsustainable under the effect of process variations, leading to unpredictable timing violations~\cite{Alam:IRPS:2011:reliability}.
For the examined circuits in this work, Fig.~\ref{fig:wc_nmed} presents the worst-case error due to timing violations (details on the experimental setup can be found in Section~\ref{sec:evaluation}).
Both the error magnitude and the peak differentiation among circuits (one order of magnitude) showcase their unpredictability and catastrophic nature.
Hence, parametric yield of a circuit (i.e., probability to meet the desired performance specification) is expected to suffer considerably under process variations and thus, reliable operation cannot be guaranteed.

Typically, parameter variability in circuits is modeled with an overly pessimistic worst-case design approach, in the form of a timing guardbands, i.e., a delay slack on top of the maximum frequency.
Extending the clock period with the inserted guardband provides the necessary leeway for tolerating degradation effects (i.e., timing violations).
However, this deliberately inserted pessimism results in significant performance degradation, which may be unacceptable for latency-critical applications~\cite{Tsiokanos:DATE:2019:prec}.
Reducing variability-induced timing guardbands can be a challenging task, especially considering that the effects of process variations are difficult to model with EDA tools via interpolation between process corners.
Furthermore, reliability-aware design methodologies often rely on different device-degrading phenomena (e.g., aging~\cite{Balaskas:TCASI:2021} and temperature~\cite{Boroujerdian:ICCD:2018:temperature}) to estimate an optimal value for timing guardbands.
Hence, variability-protecting timing guardbands are sub-optimally estimated, due to the different effect of such phenomena on CMOS devices~\cite{vSanten:IEEE-TCASI:2019:modeling}.

Therefore, it is deemed necessary to develop automated methodologies that can \emph{accurately} account for, and ultimately eliminate the catastrophic effects of process variations on digital circuits without sacrificing the application’s performance.
\emph{Approximate computing} has emerged as a promising design paradigm to solve similar degradation problems.
Approximate computing exchanges inexact computation in error-tolerant applications for gains in other design metrics (e.g., latency)~\cite{Zervakis:IEEE-Access:2020}.
Existing approximation frameworks~\cite{Faryabi:EWDTS:2015, Wang:VLSI:2017, Zhang:GLSVLSI:2020} obtain the (single) critical path (CP) with a static time analysis (STA) and then approximate it to reduce the critical path delay (CPD) of the circuit.
The premise of approximating a single path to reduce the CPD does not hold under process variations, since multiple paths can arise as critical (i.e., feature the same path delay) under different variability scenarios.
In order to obtain an actual reduction in latency, the \emph{distribution} of the CPD has to be considered.
This inefficiency (i.e., using a variability-agnostic CPD value) can be found in existing approximation frameworks that tackle circuit variability~\cite{Tsiokanos:DATE:2019:prec, Faryabi:EWDTS:2015, Wang:VLSI:2017,Zhang:GLSVLSI:2020,Gomez:TODAES:2018:dfs} and is the basis for the motivation of this work (see Section~\ref{sec:motivation} for further details).

In this work, we propose an automated variability-aware circuit approximation framework that effectively eliminates timing guardbands.
Our proposed circuit-agnostic framework operates on the gate-level netlist of any combinational circuit.
It intelligently applies netlist approximations in order to reduce both the CPD and the variance of its distribution, with minimal accuracy loss.
For that purpose, a genetic algorithm is employed to explore the vast design space of possible approximations in a fast and highly parallel manner.
The exploration procedure is aided by our high level estimators to assess the accuracy and delay of approximate solutions.
In particular, we stochastically estimate the CPD by traversing the netlist-equivalent directed acyclic graph (DAG) via a custom, variability-based version of Dijkstra's algorithm.
We build a software-based error simulator to measure the functional error of approximate circuits in a fast manner.
To accurately analyze the impact of process variations, we created $1000$ standard cell libraries, fully compatible with commercial EDA tools, for a broad range of different variation conditions.
Our framework is evaluated over several key arithmetic circuits, image processing, and Machine Learning (ML) benchmarks.
Overall, our generated variability-aware approximate circuits can be reliably operated under process degradations without timing guardbands, while a negligible accuracy loss is incurred.

\textbf{Our main contributions in this work are as follows:}
\begin{enumerate}[wide, labelwidth=!, labelindent=0pt, label=\textbf{(\arabic*)}, ref=(\arabic*)]
\item
We propose a circuit-agnostic automated framework for generating variability-aware approximate circuits.
\item
We are the first to create variability-aware standard cell libraries to accurately characterize the impact of process variations on circuits.
Our libraries are fully compatible with commercial EDA tools, greatly accelerating our framework.
\item
We propose a high-level delay estimator which stochastically traverses the netlist's DAG and models the delay distribution across multiple candidate CPs under process variations.
\item
We demonstrate that with a negligible, purely functional error (NMED) of $5.3 \times 10^{-3}$, introduced at design time, our approximate designs can deliver significant performance gains while guaranteeing the elimination of variability-induced timing violations.
\end{enumerate}

\section{Related work}\label{sec:related}
Over the past years, the mitigation of process-induced degradation on circuit performance has been the focus of significant research activities~\cite{Brendler:VLSI-SoC:2018}.
Existing works aiming to reduce the required process-induced timing guardbands employ techniques such as gate-sizing~\cite{Gomez:ET:2019:gate_sizing1, Ebrahimipour:EmTopComp:2020:gate_sizing2} or voltage scaling~\cite{Raji:Access:2021:voltage_scaling}.
A unified random telegraph noise (RTN) \& bias temperature instability (BTI) model has previously enabled the employment of probabilistic guardbands~\cite{vSanten:IEEE-TCASI:2017:unified}, to deal with joint impact of several key reliability degradations (e.g., aging, noise, etc.).
Guardband reduction in~\cite{vSanten:IEEE-TCASI:2019:modeling} was based in an estimation of the upper bound of time-dependent variability, accompanied with an automatic selection of variability-resilient cells.
Nevertheless, these methodologies are case-specific (i.e., non-automated) and may introduced design overhead.

With the emergence of approximate computing, several approximation-based methodologies have been proposed to increase circuit reliability and combat runtime degradations (e.g., aging and temperature).
An automated framework was proposed in~\cite{Balaskas:TCASI:2021} to suppress aging degradations and eliminate timing guardbands by intelligently applying approximations at the gate-level netlist.
Authors in~\cite{Amrouch:DAC:2017:towards} demonstrated the effectiveness of aging-aware approximate circuits on diminishing timing guardbands, by applying precision scaling on adders.
An approximation framework was proposed in~\cite{Kim:IEEE-TCASI:2020:aging}, where the precision of a target adder/multiplier is reduced when aging-induced timing violations on the CP are detected by a monitoring scheme.
In~\cite{Zhang:GLSVLSI:2020}, guardband reduction is achieved by iteratively applying logic approximation at the CP while performing reliability simulation to minimize path failure rate (PFR) due to aging.
In~\cite{Boroujerdian:ICCD:2018:temperature}, the authors attempted to reduce pessimistic temperature-induced timing guardbands by dynamically applying approximations, either with logic sharing or duplication of the design.
However, none of the aforementioned works directly target process variations in their attempt to reduce timing guardbands, and thus cannot guarantee reliable operation across the variability spectrum.

To that end, some works utilize approximate computing for diminishing process-induced pessimistic guardbands.
A precision scaling-based technique was proposed in~\cite{Tsiokanos:DATE:2019:prec} to detect the excitation of operands at run-time and approximate the corresponding critical paths.
Though, the proposed implementation addresses carry propagation designs (i.e., ripple carry adders).
Authors in~\cite{Faryabi:EWDTS:2015} substitute individual blocks with their approximate counterparts on paths exceeding a delay threshold, imposed by a timing guardband to simulate process-induced delay fluctuations.
However, critical paths obtained from variability-unaware STA are targeted, which can produce sub-optimal delay estimations under process variations (see Section~\ref{sec:motivation}).
Targeting the suppression of process-related timing violations of neural networks (NN), at the system level, \cite{Wang:VLSI:2017} proposed to iteratively lower the operational frequency and retrain, to adopt the tolerance of weights to timing errors.
Nevertheless, the effect of process variations at circuit level is not addressed.
In~\cite{Gomez:TODAES:2018:dfs}, an elegant solution for CP selection in the presence of aging degradations and process variation effects to enable reliable frequency scaling is proposed.
However, CPD estimations are based on stochastically-generated workload profiles and operating conditions, which mainly affect the aging rate, whereas process effects are not incorporated.

Compared to the aforementioned approaches, our framework is both automated and circuit-agnostic and directly targets process variation effects at the circuit level.
Additionally, it is also equipped to accurately measure the impact of variability-induced degradations, via specially designed standard cell libraries.
Our libraries are fully compatible with commercial EDA tools, which greatly accelerates our design flow compared to related approaches.

\section{Motivation}\label{sec:motivation}
\begin{figure}
    \centering
    \subfloat[\label{fig:motiv_example_simple_a}]{
        \includegraphics[width=0.46\columnwidth]{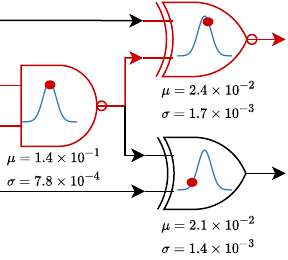}
    }
    \hfill
    \subfloat[\label{fig:motiv_example_simple_b}]{
        \includegraphics[width=0.46\columnwidth]{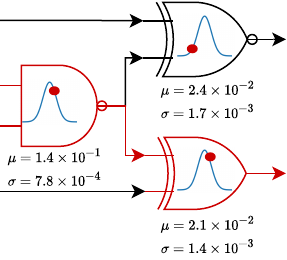}
    }
    \caption{Demonstrative motivational example of the different critical paths that may arise under process variations. Red points indicate the gate delay at random samples from each delay distribution (i.e., different process conditions). (a) A single degradation scenario may lead the critical path (CP) through the XNOR gate, whereas (b) another condition may generate a CP comprising the XOR gate.}
    \label{fig:motiv_example_simple}
\end{figure}

A key inefficiency of state-of-the-art frameworks that target variability-driven approximation is the identification of a single CP under process variations, as explained in Sections~\ref{sec:intro} and~\ref{sec:related}.
We demonstrate that this approach leads to sub-optimal results since selecting the critical path(s) to approximate is not trivial under process variations.
Fig.~\ref{fig:motiv_example_simple} provides a demonstrative motivational example for CPD estimation under process variations, for two distinct feasible situations.
Each gate is independently annotated with the characteristics (i.e., mean and standard deviation) of its delay distribution (denoted in orange), accounting for variability-related fluctuations, as obtained from Synopsys PrimeTime.
In Fig.~\ref{fig:motiv_example_simple_a}, a CP with delay $0.142ns$ includes the NAND and XNOR gates, as determined by the randomly sampled propagation delay upon each distribution (denoted by red dots).
The alternative path (i.e., NAND and XOR) falls short under the specific variability scenario, with delay $0.141ns$.
However, assuming the existence of a single CP would lead to erroneous CPD calculation for a different degradation scenario.
In Fig.~\ref{fig:motiv_example_simple_b}, degradations shift gate delays to different regions of the distributions and thus, lead to a different CP, with the NAND and XNOR gates (with CPD of $0.1417ns$, higher than the alternative of $0.1408ns$).
Thus, in Fig.~\ref{fig:motiv_example_simple}, a change in the CP is noticed at different variability scenarios.
Therefore, an approximation framework that examined only the degradations of Fig.~\ref{fig:motiv_example_simple_a} and targeted its upper (red) path, would fail to reduce the circuit's latency under the different scenario of Fig.~\ref{fig:motiv_example_simple_b}.
Hence, CPD calculation under process variations should incorporate the gate delay distributions, to effectively identify the proper approximation targets.

\begin{figure}[t]
    \centering
    \resizebox{1\columnwidth}{!}{\includegraphics{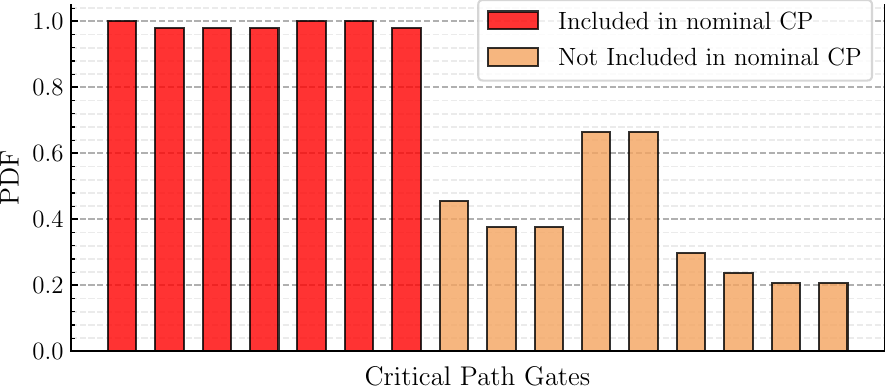}}
    \caption{
    Motivational case study for approximating multiple paths. An 8-bit adder is considered. X-axis comprises different gates while Y-axis shows their probability density function (PDF) to appear at a critical path under different variation. In red the gates that belong at the critical path of the variability-agnostic (nominal) circuit, while in orange the gates that do not.
}
    \label{fig:motiv_example}
\end{figure}

To further motivate the need for approximation frameworks with variability-aware CPD calculation,
we extend the findings of Fig.~\ref{fig:motiv_example_simple} to a motivational case study on an established circuit.
We conducted a variability Monte Carlo (MC) analysis of $1000$ points on an 8-bit adder from the industry-strength Synopsys DesignWare library~\cite{synopsys}.
Thus, we accurately determine at each condition the gates that belong to a CP (see Section~\ref{sec:libraries} for more details on the accurate representation of different process conditions with a single Monte Carlo point).
In addition, as a reference point, we evaluate the 8-bit adder without accounting for variability degradations (namely, nominal point).
Fig.~\ref{fig:motiv_example} presents the probability density function that a gate appears at a critical path.
Gates depicted in red bars belong to the CP of the circuit at the nominal point and would be targeted for approximation by traditional approximate computing frameworks~\cite{Tsiokanos:DATE:2019:prec, Faryabi:EWDTS:2015, Zhang:GLSVLSI:2020}.
However, such an approach would ignore the rest of the gates (represented by orange bars) which contribute to a considerable amount of potential critical paths under different variability degradations.
Thus, considering only gates in red would lead to sub-optimal results and the circuit's delay might not be reduced under variability effects, i.e., existing frameworks~\cite{Tsiokanos:DATE:2019:prec, Faryabi:EWDTS:2015, Wang:VLSI:2017, Zhang:GLSVLSI:2020} will fail in completely eliminating process variation-induced timing guardbands.

\section{Variability-Aware Standard Cell Libraries}\label{sec:libraries}
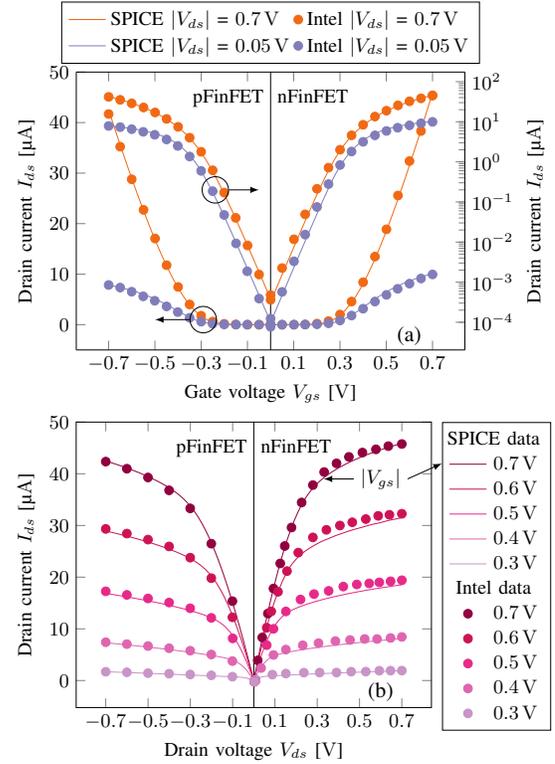
\begin{figure}
    \centering
    \small
    \resizebox{0.82\columnwidth}{!}{
    \begin{tikzpicture}
        \pgfplotsset{
            set layers = standard,
            scale only axis,
            width = 0.71\columnwidth,
            height = 4.5cm,
            cycle list = {
                {Oranges-H, smooth, mark=none},
                {Purples-H, smooth, mark=none},
                {Oranges-H, only marks, mark=*},
                {Purples-H, only marks, mark=*},
            },
        }
        \begin{axis}[
            axis y line* = left,
            xlabel = {Gate voltage \(V_{gs}\) [\si{\volt}]},
            ylabel = {Drain current \(I_{ds}\) [\si{\micro\ampere}]},
            xtick = {-0.7, -0.5, -0.3, -0.1, 0.1, 0.3, 0.5, 0.7},
            ytick = {0, 10, 20, 30, 40, 50},
            ymax = 50,
        ]
        \addplot table [x=nmos_vg_tcad_sat, y=nmos_id_tcad_sat, col sep=comma] {data/device_calibration_IdVgs.csv};
        \addplot table [x=nmos_vg_tcad_lin, y=nmos_id_tcad_lin, col sep=comma] {data/device_calibration_IdVgs.csv};
        \addplot table [x=nmos_vg_exp_sat, y=nmos_id_exp_sat, col sep=comma] {data/device_calibration_IdVgs.csv};
        \addplot table [x=nmos_vg_exp_lin, y=nmos_id_exp_lin, col sep=comma] {data/device_calibration_IdVgs.csv};
        \pgfplotsset{cycle list shift = -4}
        \addplot table [x=pmos_vg_tcad_sat, y=pmos_id_tcad_sat, col sep=comma] {data/device_calibration_IdVgs.csv};
        \addplot table [x=pmos_vg_tcad_lin, y=pmos_id_tcad_lin, col sep=comma] {data/device_calibration_IdVgs.csv};
        \addplot table [x=pmos_vg_exp_sat, y=pmos_id_exp_sat, col sep=comma] {data/device_calibration_IdVgs.csv};
        \addplot table [x=pmos_vg_exp_lin, y=pmos_id_exp_lin, col sep=comma] {data/device_calibration_IdVgs.csv};
        \draw (0, -10) -- (0, 60);
        \begin{pgfonlayer}{axis descriptions}
            \node[anchor = north west] at (0, 48) {nFinFET};
            \node[anchor = north east] at (0, 48) {pFinFET};
            \draw[-latex] (-0.35, 1) node[draw, circle, anchor = west, minimum size = 3ex] {} -- (-0.5, 1);
            \node[anchor = north] at (0.6, 1) {\normalsize (a)};
        \end{pgfonlayer}
        \end{axis}
        \begin{axis}[
            ymode = log,
            axis y line* = right,
            axis x line = none,
            ylabel = {Drain current \(I_{ds}\) [\si{\micro\ampere}]},
            ytick = {1e-4, 1e-3, 1e-2, 1e-1, 1e0, 1e1, 1e2},
            legend style = {at={(0.5, 1.03)}, anchor=south},
            legend columns = 2,
            legend transposed = true,
            legend cell align = left,
        ]
        \addplot table [x=nmos_vg_tcad_sat, y=nmos_id_tcad_sat, col sep=comma] {data/device_calibration_IdVgs.csv};
        \addplot table [x=nmos_vg_tcad_lin, y=nmos_id_tcad_lin, col sep=comma] {data/device_calibration_IdVgs.csv};
        \addplot table [x=nmos_vg_exp_sat, y=nmos_id_exp_sat, col sep=comma] {data/device_calibration_IdVgs.csv};
        \addplot table [x=nmos_vg_exp_lin, y=nmos_id_exp_lin, col sep=comma] {data/device_calibration_IdVgs.csv};
        \pgfplotsset{cycle list shift = -4}
        \addplot table [x=pmos_vg_tcad_sat, y=pmos_id_tcad_sat, col sep=comma] {data/device_calibration_IdVgs.csv};
        \addplot table [x=pmos_vg_tcad_lin, y=pmos_id_tcad_lin, col sep=comma] {data/device_calibration_IdVgs.csv};
        \addplot table [x=pmos_vg_exp_sat, y=pmos_id_exp_sat, col sep=comma] {data/device_calibration_IdVgs.csv};
        \addplot table [x=pmos_vg_exp_lin, y=pmos_id_exp_lin, col sep=comma] {data/device_calibration_IdVgs.csv};
        \addlegendentry{SPICE \(\lvert V_{ds} \rvert\) = \SI{0.7}{\volt}}
        \addlegendentry{SPICE \(\lvert V_{ds} \rvert\) = \SI{0.05}{\volt}}
        \addlegendentry{Intel \(\lvert V_{ds} \rvert\) = \SI{0.7}{\volt}}
        \addlegendentry{Intel \(\lvert V_{ds} \rvert\) = \SI{0.05}{\volt}}
        \begin{pgfonlayer}{axis descriptions}
            \draw[-latex] (-0.18, 2e-1) node[draw, circle, anchor = east, minimum size = 3ex] {} -- (-0.05, 2e-1);
        \end{pgfonlayer}
        \end{axis}
    \end{tikzpicture}%
    }
    \resizebox{0.82\columnwidth}{!}{
    \begin{tikzpicture}
        \pgfplotsset{set layers = standard}
        \begin{axis}[
            scale only axis,
            width = 0.65\columnwidth,
            height = 4.63cm,
            xlabel = {Drain voltage \(V_{ds}\) [\si{\volt}]},
            ylabel = {Drain current \(I_{ds}\) [\si{\micro\ampere}]},
            xtick = {-0.7, -0.5, -0.3, -0.1, 0.1, 0.3, 0.5, 0.7},
            ytick = {0, 10, 20, 30, 40, 50},
            ymax = 50,
            legend pos = outer north east,
            cycle list = {
                {PuRd-L, smooth, mark=none},
                {PuRd-J, smooth, mark=none},
                {PuRd-H, smooth, mark=none},
                {PuRd-G, smooth, mark=none},
                {PuRd-F, smooth, mark=none},
                {PuRd-L, only marks, mark=*},
                {PuRd-J, only marks, mark=*}, 
                {PuRd-H, only marks, mark=*},
                {PuRd-G, only marks, mark=*},
                {PuRd-F, only marks, mark=*},
            },
        ]
        \addlegendimage{empty legend}
        \addplot table [x=tcad_vd, y=tcad_id_0.7v, col sep=comma] {data/device_calibration_IdVds_nmos.csv};
        \addplot table [x=tcad_vd, y=tcad_id_0.6v, col sep=comma] {data/device_calibration_IdVds_nmos.csv};
        \addplot table [x=tcad_vd, y=tcad_id_0.5v, col sep=comma] {data/device_calibration_IdVds_nmos.csv};
        \addplot table [x=tcad_vd, y=tcad_id_0.4v, col sep=comma] {data/device_calibration_IdVds_nmos.csv};
        \addplot table [x=tcad_vd, y=tcad_id_0.3v, col sep=comma] {data/device_calibration_IdVds_nmos.csv};
        \addlegendimage{empty legend}
        \addplot table [x=exp_vd_0.7v, y=exp_id_0.7v, col sep=comma] {data/device_calibration_IdVds_nmos.csv};
        \addplot table [x=exp_vd_0.6v, y=exp_id_0.6v, col sep=comma] {data/device_calibration_IdVds_nmos.csv};
        \addplot table [x=exp_vd_0.5v, y=exp_id_0.5v, col sep=comma] {data/device_calibration_IdVds_nmos.csv};
        \addplot table [x=exp_vd_0.4v, y=exp_id_0.4v, col sep=comma] {data/device_calibration_IdVds_nmos.csv};
        \addplot table [x=exp_vd_0.3v, y=exp_id_0.3v, col sep=comma] {data/device_calibration_IdVds_nmos.csv};
        \pgfplotsset{cycle list shift = -10}
        \addplot table [x=tcad_vd, y=tcad_id_0.7v, col sep=comma] {data/device_calibration_IdVds_pmos.csv};
        \addplot table [x=tcad_vd, y=tcad_id_0.6v, col sep=comma] {data/device_calibration_IdVds_pmos.csv};
        \addplot table [x=tcad_vd, y=tcad_id_0.5v, col sep=comma] {data/device_calibration_IdVds_pmos.csv};
        \addplot table [x=tcad_vd, y=tcad_id_0.4v, col sep=comma] {data/device_calibration_IdVds_pmos.csv};
        \addplot table [x=tcad_vd, y=tcad_id_0.3v, col sep=comma] {data/device_calibration_IdVds_pmos.csv};
        \addplot table [x=exp_vd_0.7v, y=exp_id_0.7v, col sep=comma] {data/device_calibration_IdVds_pmos.csv};
        \addplot table [x=exp_vd_0.6v, y=exp_id_0.6v, col sep=comma] {data/device_calibration_IdVds_pmos.csv};
        \addplot table [x=exp_vd_0.5v, y=exp_id_0.5v, col sep=comma] {data/device_calibration_IdVds_pmos.csv};
        \addplot table [x=exp_vd_0.4v, y=exp_id_0.4v, col sep=comma] {data/device_calibration_IdVds_pmos.csv};
        \addplot table [x=exp_vd_0.3v, y=exp_id_0.3v, col sep=comma] {data/device_calibration_IdVds_pmos.csv};
        \addlegendentry{\hspace{-5ex}SPICE data}
        \addlegendentry{\SI{0.7}{\volt}}
        \addlegendentry{\SI{0.6}{\volt}}
        \addlegendentry{\SI{0.5}{\volt}}
        \addlegendentry{\SI{0.4}{\volt}}
        \addlegendentry{\SI{0.3}{\volt}}
        \addlegendentry{\hspace{-5ex}Intel data}
        \addlegendentry{\SI{0.7}{\volt}}
        \addlegendentry{\SI{0.6}{\volt}}
        \addlegendentry{\SI{0.5}{\volt}}
        \addlegendentry{\SI{0.4}{\volt}}
        \addlegendentry{\SI{0.3}{\volt}}
        \draw (0, -10) -- (0, 60);
        \begin{pgfonlayer}{axis descriptions}
            \node[anchor = north west] at (0, 48) {nFinFET};
            \node[anchor = north east] at (0, 48) {pFinFET};
            \node[anchor = north] at (0.6, 1) {\normalsize (b)};
            \node (vgs) at (0.6, 39) {\(\lvert V_{gs} \rvert\)};
            \draw[-latex] (vgs.west) -- (0.33, 39);
            \draw[-latex] (vgs.east) -- (0.89, 42);
        \end{pgfonlayer}
        \end{axis}
    \end{tikzpicture}%
    }
    \caption{The industry compact model for FinFET (BSIM-CMG) is carefully calibrated to reproduce Intel \SI{14}{\nano\metre} measurement data extracted from~\cite{intel_data}.
    SPICE simulations (using our calibrated models) achieve an excellent agreement with the measurement data for both nFinFET and pFinFET devices.
    The top figure~(a) shows the validation for the case of $I_{ds}$-$V_{gs}$ at high and low $V_{ds}$ biases.
    The bottom figure~(b) shows the validation for the case of $I_{ds}$-$V_{ds}$ at various $V_{gs}$ biases.
    Figure data were obtained from~\cite{FinFET_var}.}
    \label{fig:device_calibration}
\end{figure}

\begin{figure}
    \centering
    \small
    \resizebox{0.85\columnwidth}{!}{
    \begin{tikzpicture}
        \pgfplotsset{
            set layers = standard,
            scale only axis,
            width = 0.43\columnwidth,
            height = 3.5cm,
        }
        \begin{groupplot}[
            group style={
                group name = my plots,
                group size = 2 by 1,
                horizontal sep = 0cm,
                ylabels at = edge left,
                yticklabels at = edge left,
            },
            xlabel = {\(I_\text{on}\) [\si{\milli\ampere\per\nano\metre}]},
            ylabel = {\(I_\text{off}\) [\si{\nano\ampere\per\nano\metre}]},
            ymode = log,
            ymin = 1,
            ymax = 1000,
            ymajorgrids,
        ]
        \nextgroupplot [
            xmin = 0.7,
            xmax = 1.399,
            xtick = {0.8, 0.9, 1.0, 1.1, 1.2, 1.3},
        ]
        \begin{pgfonlayer}{axis descriptions}
            \node [anchor = west] at (rel axis cs: 0.1, 0.5) {nFinFET};
        \end{pgfonlayer}
        \nextgroupplot [
            xmin = 0.601,
            xmax = 1.3,
            xtick = {0.7, 0.8, 0.9, 1.0, 1.1, 1.2},
        ]
        \begin{pgfonlayer}{axis descriptions}
            \node [anchor = west] at (rel axis cs: 0.1, 0.5) {pFinFET};
        \end{pgfonlayer}
        \end{groupplot}
        \begin{groupplot}[
            group style={
                group name = my plots,
                group size = 2 by 1,
                horizontal sep = 0cm,
                ylabels at = edge left,
                yticklabels at = edge left,
            },
            ticks = none,
            legend pos = north west,
            legend columns = 2,
        ]
        \nextgroupplot [
            xmin = 320.6,
            xmax = 1440.669,
            ymin = 221.526,
            ymax = 1200.495,
        ]
        \addplot [thick, black] table [x index=0, y index=1, col sep=comma] {data/variability_calibration/nmos_black.csv};
        \addplot [thick, dashed, Reds-J] table [x index=0, y index=1, col sep=comma] {data/variability_calibration/nmos_red.csv};
        \addplot [only marks, mark=o, Blues-J, opacity=0.5] table [x index=0, y index=1, col sep=comma] {data/variability_calibration/nmos_circles.csv};
        \legend{Regression line (SPICE), Regression line (Intel), Monte-Carlo simulations}
        \nextgroupplot [
            xmin = 1820.54,
            xmax = 2897.2599999999998,
            ymin = 221.526,
            ymax = 1170.495,
        ]
        \addplot [thick, black] table [x index=0, y index=1, col sep=comma] {data/variability_calibration/pmos_black.csv};
        \addplot [thick, dashed, Reds-J] table [x index=0, y index=1, col sep=comma] {data/variability_calibration/pmos_red.csv};
        \addplot [only marks, mark=o, Blues-J, opacity=0.5] table [x index=0, y index=1, col sep=comma] {data/variability_calibration/pmos_circles.csv};
        \end{groupplot}
    \end{tikzpicture}%
    }
    \caption{Variability calibration of our FinFET compact model against Intel \SI{14}{\nano\metre} measurement data~\cite{intel_data}. The regression line obtained from Monte-Carlo SPICE simulations on the transistor model is in good agreement with the data from Intel variability measurements.
    Figure data were obtained from~\cite{FinFET_var}.}
    \label{fig:variability_calibration}
\end{figure}
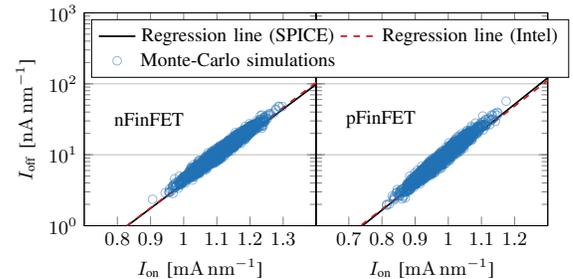

\textbf{Technology Modeling and Calibration:}
In our work, we employ the \SI{14}{\nano\metre} FinFET technology, that is fully calibrated to reproduce FinFET measurements from Intel, extracted from~\cite{intel_data}. 
The measurement data from Intel are based on a mature high-volume manufacturing process.
In \figurename{}~\ref{fig:device_calibration}, we demonstrate the calibration and benchmarking against measurements for both n-type and p-type FinFET transistors for a wide range of $I_D$-$V_G$ and $I_D$-$V_D$ biases.
Further, the variability that FinFET exhibits in the examined \SI{14}{\nano\metre} technology node has been also carefully calibrated and benchmarked against measurement data from Intel for the same technology node, as depicted in \figurename{}~\ref{fig:variability_calibration}.
The considered underlying sources of manufacturing variation are the metal gate work-function ($\phi_g$), channel length ($L_g$), fin height ($H_{fin}$), fin thickness ($T_{fin}$) and effective oxide thickness ($EOT$).
The industry-standard compact model of FinFET technology (BSIM-CMG)~\cite{BSIM-CMG} is calibrated through iteratively tuning the transistor model-card parameters. 
The main parameters are low-field mobility, sub-threshold swing ($SS$), velocity saturation, drain induced barrier lowering, and series resistance in which SPICE simulations reproduce Intel's measurements. 
Afterwards, using Monte-Carlo simulations along with the above-mentioned calibrated industry compact model, the standard deviation of studied variability sources are carefully calibrated to match the reported variability measurement data of Intel \SI{14}{\nano\metre} FinFET. 
Further details on our transistor and variability calibrations are available in~\cite{FinFET_var}. 
In that work, we have demonstrated that SPICE simulations for both nFinFET and pFinFET transistors have an excellent match with the measurement data and this holds for the various $I_D$-$V_G$ and $I_D$-$V_D$ biases. 
In addition, we have also demonstrated that the variation in the $I_{ON}$ vs. $I_{OFF}$, obtained from our SPICE Monte-Carlo simulations comes in excellent agreement with the Intel measurement data for both nFinFET and pFinFET.

\begin{figure}[t]
    \centering
    \includegraphics[width=0.9\columnwidth]{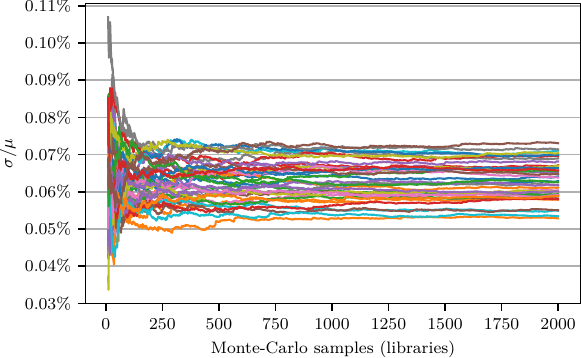}
    \caption{
    Variations in cell timing information as a function of the number of Monte-Carlo cell libraries.
    Each curve in the graph represents the signal propagation delay of an AND-gate from our cell libraries (6~differently sized AND-gates in total).
    The relative standard deviation ($\sigma/\mu$) of the propagation delays among all cell libraries heavily fluctuates for a few amount of samples ($500$). 
    However, with more samples being added to our collection of cell libraries, the relative standard deviation continuously converges.
    At the $1000$ sample mark, the variations have stabilized and thus, the variability of all devices is well reflected.
    }
    \label{fig:rev:mc_samples_plot}
\end{figure}

\textbf{Variability-Aware Cell Library Characterization:} 
As explained above, the calibrated compact models capture both the electrical characteristics of transistors as well as variation effects. 
To perform standard cell library characterization, we employ calibrated compact models along with the post-layout SPICE netlists of standard cells. 
The latter was obtained from the open-source NanGate FinFET-based standard cell library~\cite{nangate15}, which is publicly available at~\cite{Silvaco}. 
Our cell library characterization was performed using Synopsys tool flows~\cite{synopsys}. 
It is noteworthy that every standard cell was characterized under $7 \times 7$ operating conditions (i.e.,~input signal slew and output load capacitance).
During the characterization process, both delay and power (static and switching power) were measured by HSPICE and then stored using the standard ``Liberty'' format to form the created standard cell library.
In practice, $1000$ standard cell libraries were characterized based on the statistical information of the underlying variability.
Finally, we analyze the delay information for every standard cell (obtained from $1000$ iterations) to calculate the corresponding mean ($\mu$) and standard deviation ($\sigma$).
The amount of libraries (i.e., $1000$) was carefully selected via experimental evaluation to balance the time complexity of our framework with a reliable assessment of device variability.
Fig.~\ref{fig:rev:mc_samples_plot} depicts the relative standard deviation ($\sigma/\mu$) of cell delay per amount of cell libraries used for its calculation, via Monte-Carlo analysis.
As observed, a small sub-sample of libraries ($<500$) leads to high fluctuations of the $\sigma/\mu$ ratio.
At the $1000$-library mark, these values have stabilized enough to acquire a reliable measurement which accurately reflects the device variability.
We obtain for every standard cell twice the number of $7 \times 7$ look-up-tables for timing information; one for $\mu$ and one for $\sigma$.
This enables us to store the statistical information of the entire standard cell library using the standard ``Liberty Variation Format'' (LVF).

Importantly, all our created $1000$ standard cell libraries are fully compatible with the existing commercial EDA tool flows (e.g., Cadence and Synopsys). 
Hence, we can directly deploy them to perform Static Timing Analysis (STA) to extract the delay of paths in a circuit despite the complexity of the circuit and/or the path. 
Similarly, our additional LVF-based standard cell library is also compatible with the existing commercial EDA tool flows.
Hence, we can directly deploy it to perform Statistical Static Timing Analysis (SSTA) and extract both the $\mu$ and $\sigma$ of any path within a circuit, despite the complexity.
In our work, we employ Synopsys STA and SSTA tool flows to perform the required timing and statistical analysis of circuits (details in Section~\ref{sec:methodology}). 

Compared to the existing state of the art, in our work, we rely on a pure Monte-Carlo approach to characterize the standard cell libraries.
The latter allows us to accurately evaluate the impact of variability on any path within a circuit using STA.
This is different from the existing work (e.g.,~\cite{FinFET_var}), in which the standard cell library is characterized under the effects of variability (e.g.,~using the so-called ``sensitivity-based'' approach~\cite{Siliconsmart}) towards directly obtaining the LVF-based cell library.
In such an approach, only one library (i.e.,~LVF-based library) is generated.

\section{Our Proposed Framework}\label{sec:methodology}
\begin{figure}[!t]
    \centering
    \resizebox{1\columnwidth}{!}{\includegraphics{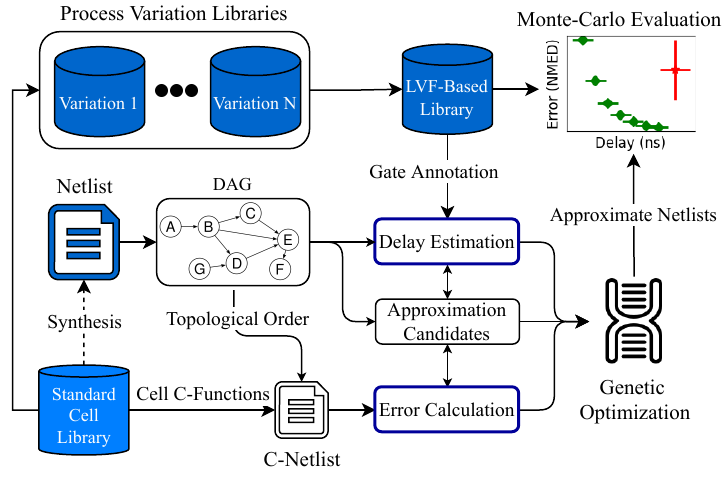}}
    \caption{Abstract overview of our proposed framework. The input of our framework is a post-synthesis gate-level netlist while its output is a Pareto-front of variabilty-aware approximate circuits, via Monte-Carlo evaluation.
    }
    \label{fig:methodology}
\end{figure}

In this work, we propose an automated variability-aware circuit approximation framework that efficiently eliminates variability-induced timing guardbands.
The abstract overview of our proposed framework is presented in Fig.~\ref{fig:methodology}.
Our framework operates at the gate-level netlist of arbitrary circuits, if it is available, otherwise standard EDA synthesis tools (e.g., Synopsys) can be used to generate it.
Exploiting the regular structure of gate-level netlists is crucial for automating our framework and making it circuit-agnostic.

The goal of our optimization is to generate variability-aware approximate circuits that do not exhibit timing violations under process variations, thus eliminating the need to employ any timing guardbands.
We reduce the circuit CPD along with its variability (i.e., how much the CPD variates under different variability-induced degradations) to fulfill this goal.
For that purpose, we employ approximate computing principles.
Specifically, we leverage the truncating nature of netlist approximations to cut-off potentially critical paths and as a result, accelerate the design (i.e., reduce its CPD).
During optimization, approximations are selectively applied at the node level of the netlist-equivalent DAG to generate approximate circuits of reduced latency.
We develop a multi-objective genetic algorithm (GA) to efficiently explore the design space of possible approximate circuits and derive to a close-to-Pareto-optimal set of solutions with diminished delay variability and minimal functional error.
The fast and parallel execution of the GA is enabled by our high level estimators, for functional error calculation and CPD estimation.
Finally, we evaluate the obtained approximate circuits through a 1000-point Monte-Carlo analysis, using the aforementioned variation libraries, to obtain their delay characteristics (see Section~\ref{sec:libraries}).
The resulting Pareto front consists only of circuits without timing errors and with functional NMED lower than the worst-case baseline NMED.

\subsection{Circuit Netlist Approximations}\label{sec:approximations}

\begin{figure}
    \centering
    \resizebox{1\columnwidth}{!}{\includegraphics{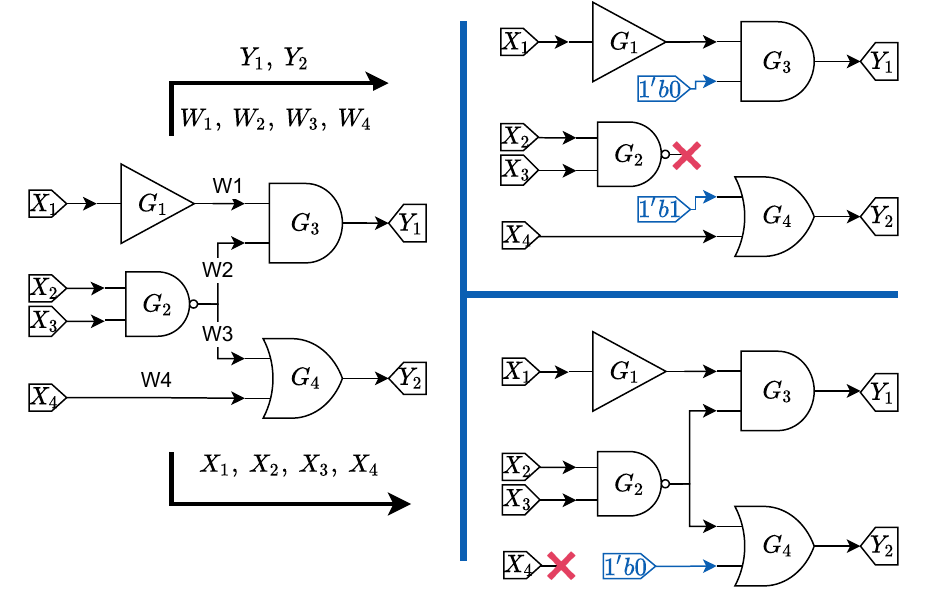}}
    \caption{
    Gate-level netlist approximation techniques, implemented in our framework. On the baseline circuit (left), gate-level pruning (top right) and precision scaling (bottom right) are applied.
    }
    \label{fig:ax_techniques}
\end{figure}

We utilize gate-level approximation as a mechanism for accelerating selected paths within the netlist.
In other words, we induce approximations which alter the structure of the paths comprised in the netlist or the connectivity between its gates with the aim of reducing the circuit CPD.
The approximation techniques employed synergistically in our framework are gate-level pruning (GLP)~\cite{Schlachter:IEEE-VLSI:2017:glp} and precision scaling~\cite{Amrouch:DAC:2017:towards}.
The former involves substituting a wire by a constant value (i.e., 0 or 1), thus pruning the gate that drives it.
The path comprising that wire becomes potentially faster as the signal propagation is cut off at the point of substitution.
Precision scaling~\cite{Amrouch:DAC:2017:towards} is an algorithmic approximation technique which involves the selective truncation of the least significant input bits (LSB), i.e., setting bits of the circuit's inputs to zero.
Both techniques utilized in our framework are presented in Fig.~\ref{fig:ax_techniques}.
We exploit the truncating nature of precision scaling to express it as a specific case of GLP, where an input wire is still substituted by a constant value (see the top right of Fig.~\ref{fig:ax_techniques}), without though pruning a netlist gate.
Thus, both techniques are seamlessly combined and implemented at gate-level, by manipulating the connections between nodes (gates) in the DAG (netlist), as can be seen in Fig.~\ref{fig:ax_techniques}.
The integration of wire approximations at the DAG allows us to avoid certain inefficiencies, traditionally associated with the aforementioned techniques.
Originally, the GLP technique selects approximation candidates (i.e., gates to be pruned) with a significance and activity criterion~\cite{Schlachter:IEEE-VLSI:2017:glp}.
This however severely limits the design space and consequently, the possible performance gains and error compensation possibilities.
Additionally, precision scaling is typically conducted in an iterative, greedy manner to discover which combination of input bits to truncate is less harmful to the output accuracy, which is computationally inefficient.

It is noteworthy that our framework is not confined to the specific approximation techniques.
In fact, other netlist approximation techniques (e.g., wire-by-wire substitution~\cite{Jain:DATE:2016:wbw}) are orthogonal to our method and can be analogously applied in unison for further performance gains.

\subsection{High-Level Accuracy and Delay Estimators}\label{sec:tools}
To facilitate the timely convergence and scalability of our framework, we develop high-level estimators for appraising circuit delay and accuracy during optimization.
The DAG representation can directly model inter-gate dependencies within timing paths, which can be exploited to gather the critical paths under variability effects.
Additionally, we build a netlist software-based (i.e., untied from gate delay information) error simulator at high level to avoid lengthy circuit simulations by EDA tools.

\subsubsection{Error Estimation with Software-Based Simulation}\label{sec:csim}
We build a software-based error simulator to measure the functional error of our approximate circuits in a fast and scalable manner, during optimization.
Typically, circuit simulations are used to accurately calculate the circuit error.
Conducting them within the optimization flow however would be very time consuming.
Moreover, circuit simulations would diminish the scalability of our framework, since commercial simulation tools (e.g., QuestaSim) require licensing agreements, thus limiting any potential parallelization.
To avoid this caveat, we build a software-based error simulator at high level, similar to the one described in~\cite{Zervakis:IEEE-Access:2020}, based on the the assumption that timing errors are eventually eliminated and the only source of error is our functional approximation.
Simulations are conducted using C and are integrated into our Python-built optimization procedure using ctypes.
To build the C-equivalent of the gate-level netlist, we translate the Verilog description of the standard cell library into C (one-time effort) and model the cells as C-functions.
Since Verilog code is concurrent and C is sequential, the topological ordering of the DAG ensures the correct distillation of the circuit into a C function.
The DAG edges (wires) serve as inputs and outputs of the C function.
Thus, we can accurately perform functional error simulations with high precision at high level for every approximate circuit generated by our optimization, in a fast and highly parallelizable manner using large datasets.

As an accuracy metric to quantify both the rate and the magnitude of the introduced approximation errors, we selected the normalized mean error distance (NMED)~\cite{Momeni:TC:2014:NMED}:
\begin{align}\label{eq:med}
NMED = \frac{1}{max}\frac{1}{N}\sum_{i=0}^N \frac{|Y-\hat{Y}|}{Y},
\end{align}
where $Y$ is the accurate output, $\hat{Y}$ the approximate output, $N$ the number of inputs and $max$ is the maximum possible absolute error value.

\subsubsection{Delay Estimation with Stochastic DAG Traversal}\label{sec:delay}
Estimating the CPD  under process variations is crucial for accelerating our optimization methodology in a scalable manner.
We build a delay estimator that stochastically traverses the DAG to gather its potential critical paths under variability-induced degradations and probabilistically estimate the circuit's CPD.

As mentioned in Section~\ref{sec:intro}, a common inefficiency in typical approximation frameworks is conducting variability-unaware STA when selecting the paths to approximate.
To achieve accurate estimations, we run an SSTA on the input gate-level netlist with our variability-aware standard cell libraries in their LVF-based format.
Thus, we annotate each gate with a  a nominal delay value and the corresponding standard deviation, creating a statistical profile of its delay variability.
Gate annotations are dependent to the type of the corresponding standard cell, which allows us to model them as independent random variables (RV), i.e., normally distributed variables.
In detail, the statistical profile of each RV includes multiple timing transitions (e.g., rising/falling delay, with negative/positive unate), as provided by the EDA STA tool.
Note, while state-of-the-art-approaches might statistically model the delay fluctuations due to process variations, we accurately annotate them using the LVF data of our variability-aware libraries.

Here, we explain in detail the stochastic DAG traversal.
We can mathematically formulate the DAG traversal using the principles of Dijkstra's algorithm, but instead of using a numerical value to represent a gate's propagation delay (as would be the case when only considering the nominal delay) we use RVs.
The mathematical formulas for accumulating and comparing RVs (i.e., delay distribution of gates) within a DAG path are described in~\cite{Olya:Inderscience:2014:dijkstra}, and are leveraged in our work to create our stochastic delay estimation tool.
Similar to the vanilla Dijkstra algorithm, a path's delay distribution can be modeled as the accumulation of its comprising random variables. 
Let $X$ and $Y$ be two independent random variables with density functions $f_X(x)$ and $f_Y(y)$, $\forall x,y$.
According to~\cite{Olya:Inderscience:2014:dijkstra}, the sum $Z = X + Y$ is a random variable with density function $f_Z(z)$, where $f_Z$ is the convolution of $f_X$ and $f_Y$:

\begin{equation}\label{eq:sum_rv_densities}
    \begin{aligned}
    f_Z(z) = &\int_{-\infty}^{+\infty} f_{Y}(z-x) f_{X}(x)\,dx \\
    = & \int_{-\infty}^{+\infty} f_{X}(z-y) f_{Y}(y)\,dy.
    \end{aligned}
\end{equation}
\eqref{eq:sum_rv_densities} leads to the following finding~\cite{Olya:Inderscience:2014:dijkstra}: if $X_1$,~…,~$X_n$ are independent random variables with ($\mu_1,\sigma_1^2$),~...,~($\mu_n,\sigma_n^2$), then $Y$ follows a normal distribution as well:

\begin{equation}\label{eq:sum_rv}
\begin{aligned}
Y=\sum_{i=1}^n X_i=N\left( \sum_{i=1}^{n} \mu_i, \sum_{i=1}^{n} \sigma_i^2 \right).
\end{aligned}
\end{equation}
We utilize this property to calculate the accumulated delay of a single path in the DAG, from an input to an output node.

Comparisons occur between the RVs of consecutive nodes (i.e., connected nodes) during the traversal, to trace the potentially slowest paths.
This process is equivalent to comparing deterministic arc lengths for shortest path calculation in the vanilla Dijkstra's algorithm.
Comparing two random variables can be done by assessing the probability that one will be greater than the other:

\begin{equation}\label{eq:diff_rv}
\begin{aligned}
\mathds{P}(X\!>\!Y) \!=\! \mathds{P}(X\!-\!Y\!>\!0) \!=\! 1\!-\!\Phi\!\left( \!\frac{\E(Y) \!-\! \E(X)}{\sqrt{\V(X) \!+\! \V(Y)}} \right),
\end{aligned}
\end{equation}
where $\E$ and $\V$ denote the mean ($\mu$) and variance ($\sigma^2$) of a random variable, respectively and $\Phi$ denotes the cumulative distribution function of the normal distribution.

\begin{figure}[!t]
    \centering
    \subfloat[]{
    \includegraphics[width=\columnwidth]{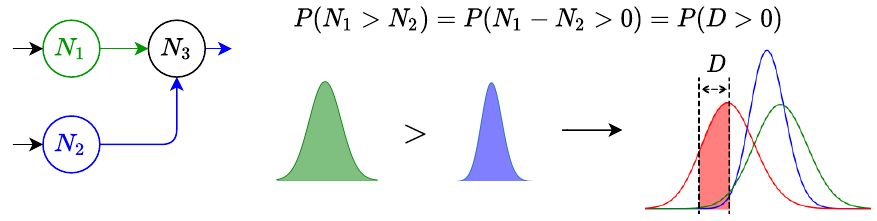}
    \label{fig:rv1}
    }
    
    \subfloat[]{
    \includegraphics[width=\columnwidth]{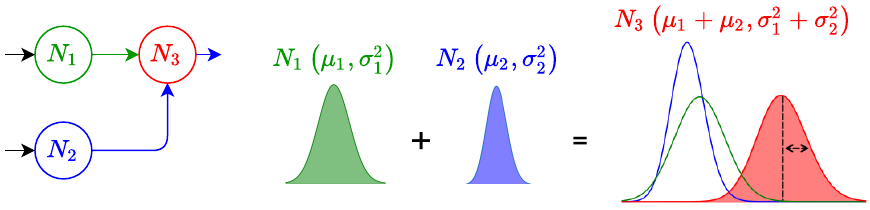}
    \label{fig:rv2}
    }
    \caption{Descriptive example of the stochastic DAG traversal utilized in our delay estimation. (a) The delay distribution of node $N_1$ is compared to the one of node $N_2$, as defined by~\eqref{eq:diff_rv}. We check if the distribution of their difference (in red) has a positive mean (the mean of their graphical overlap $D$ is above zero). (b) If it is true, then the delay distribution of $N_3$ is updated as the sum (red) of the distributions $N_2$ (blue) and $N_1$ (green), as defined by~\eqref{eq:sum_rv}.}
    \label{fig:rv}
\end{figure}

The two key actions of the stochastic DAG traversal of \eqref{eq:sum_rv}-\eqref{eq:diff_rv} are demonstrated in Fig.~\ref{fig:rv}.
By comparing (Fig.~\ref{fig:rv1}) and accumulating (Fig.~\ref{fig:rv2}) the density delay distributions of each node using \eqref{eq:diff_rv} and \eqref{eq:sum_rv}, respectively, each output node is annotated with the delay distribution of its longest (slowest) path, along with the probability that this path is longer (slower) than any other path in the DAG.
The output node featuring the highest probability is appointed as the endpoint of the critical path, and its delay distribution as the CPD.
Additionally, we consider the calculated probability as a confidence metric that under process variations, the specific path will hold as critical.
We later leverage this confidence level to encourage approximations at highly-probable critical paths.

Multiple timing transitions are incorporated within the DAG traversal, by exploiting prior knowledge from the EDA STA tool.
Specifically, we annotate each DAG edge with the transition most frequently appearing in $1000$ critical paths, as obtained from the $1000$-point Monte-Carlo evaluation of the baseline circuit (more in Section~\ref{sec:evaluation}).
This process is key for creating an accurate estimator of high correlation to the true circuit delay, since information from the STA tool is used.

Estimating the CPD with the above methodology drastically accelerates our optimization methodology and enhances its scalability.
The most accurate alternative, i.e., conducting STA with the process variation libraries for every candidate approximate circuit, would lead to unacceptably large convergence time of our optimization procedure, for two reasons:
a) STA is much slower than our delay estimation (e.g., $4.4\times$ slower for a single delay estimation on a 16-bit multiplier)
and b) STA tools are restricted by the availability of licenses, thus severely limiting parallelization.

\subsection{Approximation Candidates}\label{sec:candidates}
We present the methodology for extracting approximation candidates from the netlist, which will be next used in our optimization phase to generate the approximate circuits.
Approximation candidates are formed as a set of integers (which we call chromosome from further on), in order to be easily manipulated during optimization.
The set of possible approximation candidates $AC$ includes all wires comprising the netlist (both circuit inputs and gate output wires, according to precision scaling and GLP, respectively)  which can be substituted by either $0$ or $1$, or not approximated ($-1$):
\begin{equation}\label{eq:candidates}
\begin{gathered}
AC = \{(w,k): k \in \{-1,0,1\},\, \forall w \in W \}
\end{gathered}
\end{equation}
where $W$ denotes the set of wires in the gate-level netlist.
Originally, the design space comprises of $|W|^3$ approximate solutions (i.e., each wire can either be approximated by $0$, $1$ or remain accurate), according to~\eqref{eq:candidates}.
Its size can prove intractable for larger data paths, as execution time will be significantly slowed down, due to the prolonged DAG traversals for each approximate circuit.
Thus, to prune the design space, we limit our approximation candidates to the wires which can provide actual latency reduction, i.e., have a high chance of participating in a critical path.
This is achieved via our stochastic delay estimation model, at the DAG representation (see Section~\ref{sec:delay}).
As aforementioned, each output node is appointed a probability of constituting the terminal of a CP (let's call this probability CPB, i.e., critical path probability).
We back-propagate this probability using a simple rule from probability theory: the probability of a node to participate in a CP is the sum of the same probabilities of all of its children nodes.
For example, if the CPB of two children nodes are $0.1$ and $0.15$, then the CPB of their parent node (i.e., the connected node higher in the tree structure) is $0.25$.
Thus, we acquire a probability estimation for each node to reside in a potentially critical path.
By constraining the CPB of each node to a reasonable threshold $\text{CPB}_t$, we arrive at a pruned and more efficient (i.e., easy to traverse) design space, formulated as:
\begin{equation}\label{eq:candidates_cpb}
\begin{aligned}
AC = \{&(w,k): k \in \{0,1\},\, \\
&\forall w \in W \ | \ \text{CPB}_{Nw} \ge \text{CPB}_t \},
\end{aligned}
\end{equation}
where $\text{CPB}_{Nw}$ denotes the CPB of node $N$, whose output wire is $w$.
This design space pruning scheme allows for better exploration of \emph{meaningful} approximate solutions and consequently, faster convergence of our optimization methodology.

\begin{figure*}[t]
    \centering
    \resizebox{1\textwidth}{!}{
        \includegraphics{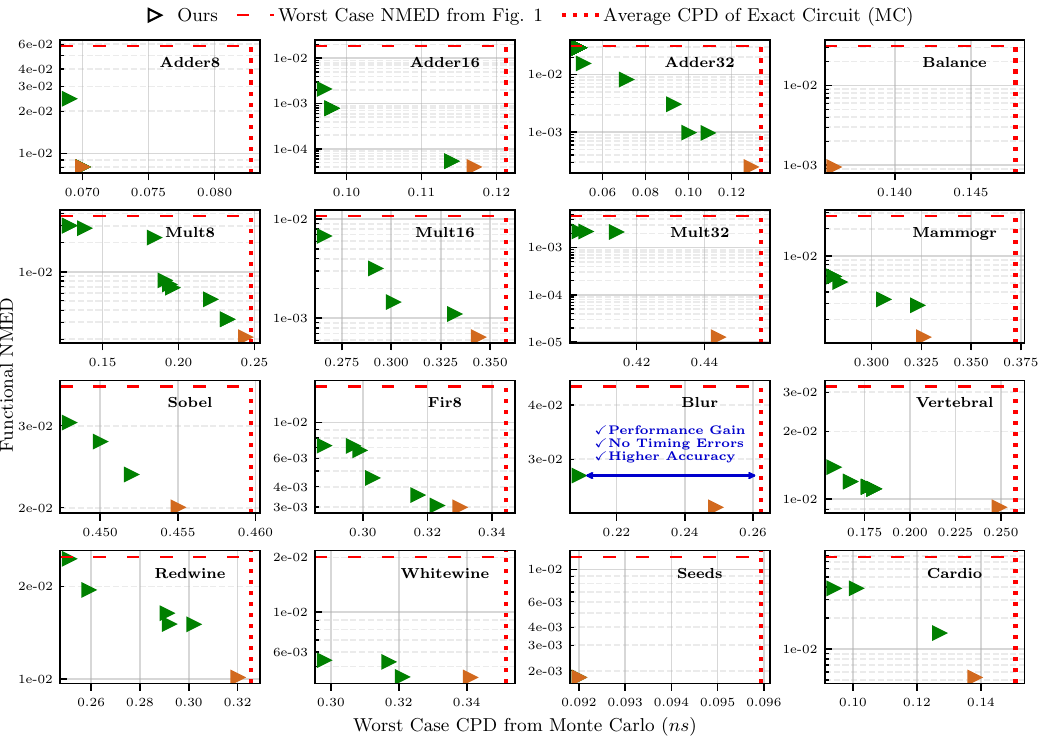}
    }
    \caption{Pareto front of our approximate solutions of all examined circuits for the functional NMED, introduced via approximation, vs the worst-case CPD, as obtained from a Monte-Carlo evaluation for each solution (i.e., data point) with $1000$ variability-aware standard cell libraries.}
    \label{fig:pareto}
\end{figure*}

\subsection{Optimization Procedure}\label{sec:genetic}
To explore the design space of possible variability-aware approximate circuits, we employ the multi-objective Non-dominated Sorting Genetic Algorithm (NSGA-II), proposed in \cite{Deb:IEEE:2002:NSGAII}.
As seen in Fig.~\ref{fig:methodology}, the GA receives the approximation candidates to build approximate solutions and our estimators to evaluate them.
The inherently parallel nature of genetic algorithms, along with our high level delay and error estimators, allows for a fast exploration of the vast design space, while exploiting the available computational resources to their full capabilities.

In an attempt to incentivize the exploration of solutions with moderate logic approximation at the initial stages of evolution, we initially create semi-random chromosomes, more inclined towards non-approximated wires.
We explicitly target three objective metrics for joint minimization: a) error (NMED), b) critical path delay (CPD), and c) standard deviation of the CPD (see Section~\ref{sec:delay}).
Thus, the obtained approximate circuits will exhibit the most dominant combination of delay variability reduction and low error among all those explored.
Additionally, we penalize approximate solutions with limited confidence regarding their CPD calculation, as given by our delay estimation methodology.
Uncertainty in the CPD calculation indicates that the delay distribution of multiple paths in the DAG resemble each other, meaning that approximating one (i.e., identifying a single CP) might not lead to the desired delay reduction across the variability spectrum.
An upper bound to the acceptable error of generated solutions is also applied: we constrain NMED to the maximum baseline error (see Fig.~\ref{fig:wc_nmed}), to discourage the exploration of solutions with unacceptably high error (despite their reduced delay).
Finally, we apply random mutation to the generated chromosomes, weighted inversely proportional to the depth of each wire in the DAG (i.e., a gene which represents a wire closer to the output would be less likely to be mutated).
Random mutations (i.e., inserted approximations) to wires close to the circuit output have a higher possibility of cutting-off all paths leading to an output bit, thus setting a constant (and potentially unrecoverable) error.
This can be particularly harmful for paths leading to the most significant bits.

The genetic algorithm outputs a close-to-Pareto-optimal set of approximate solutions w.r.t. the aforementioned objectives.
These approximate circuits are then evaluated through a Monte-Carlo simulation using our variability-aware libraries and we obtain their true delay characteristics.

\section{Results and Evaluation}\label{sec:evaluation}
In this section, we evaluate the efficiency of our automated framework in eliminating the variability-induced timing violations, thus completely removing timing guardbands due to process variations.
Several key arithmetic circuits, image processing benchmarks, and machine learning classifiers are examined.
Specifically, we target three delay-optimized adders and multipliers of 8, 16 and 32 bits, generated from the Synopsys DesignWare library.
Additionally, our framework is evaluated over more complex dataflows, built upon optimized components of the DesignWare library as well.
Three image processing benchmarks are targeted: a Sobel edge detector~\cite{Zervakis:TVLSI:2019:vader}, an 8-tap FIR filter~\cite{Paim:TCASI:2021}, and a Gaussian Blur filter~\cite{Lee:Springer2019:ahls}.
We also examine a variety of ML-related circuits. 
Three multi-layer perceptron (MLP) circuits are used, trained on the Balance, Mammographic (Mammogr), and Vertebral datasets, respectively, of the UCI ML repository~\cite{Dua:2019:datasets}.
We also consider two support vector machine (SVM) classifiers, trained on the RedWine and WhiteWine datasets, as well as two decision trees, of the Seeds and Cardiotography (Cardio) datasets, from the same repository.
Hence, our evaluation covers a wide spectrum of DSP, image processing and ML domains, i.e., perfect candidates for approximate computing.

Our variability standard cell libraries are used to conduct a $1000$-point Monte-Carlo variation-aware analysis across different process conditions.

All circuits, baseline or approximate, are synthesized with zero slack (i.e., targeting maximum performance).
Synopsys Design Compiler is used for synthesis, along with the \texttt{compile\_ultra} command for extensive netlist optimizations.
Synopsys PrimeTime is used to gather the delay distribution statistics of the generated approximate solutions by exploiting its Parametric On-Chip Variation (POCV) features.
This is crucial for annotating the DAG with information about gate delay distributions under process variations.
Moreover, PrimeTime is used to run STA and obtain the post-synthesis CPD during the Monte-Carlo analysis. 
Gate-level timing simulations are performed using Mentor Questasim to calculate a circuit's output error (NMED).
Two different input datasets of $10^5$ random, uniformly distributed inputs per circuit, are used in our framework, for the functional simulations during optimization (see Section~\ref{sec:csim}) and the baseline evaluation.
Note that using the above described experimental setup we evaluated the error of the baseline circuit (results in Fig.~\ref{fig:wc_nmed}).
In Fig.~\ref{fig:wc_nmed}, a $1000$-point Monte-Carlo is performed in which the baseline circuits are simulated at their nominal CPD (obtained with our LVF-based library, details in Section~\ref{sec:libraries}).

\subsection{Eliminating Variability-Induced Timing Violations}\label{sec:eval_mc}

\begin{table}[t]
\caption{Evaluation of selected approximate solutions from our Pareto fronts with minimum functional NMED, for all examined circuits. Reduction w.r.t. the baseline circuits is considered.}
\centering
{\footnotesize
\setlength{\tabcolsep}{4pt}
\begin{tabular}{c|c|c|c}
	\hline

	\textbf{Circuit} & \makecell{\textbf{Functional} \textbf{NMED}} & \makecell{\textbf{CPD}  \textbf{Red. (\%)}} & \makecell{\textbf{Std.} \textbf{Red. (\%)}}\\
	\hline

	Adder8 & $8.0\times10^{-3}$ & $26.4$ & $28.8$\\
	\hline
	Adder16 & $4.0\times10^{-5}$ & $15.0$ & $15.7$\\
	\hline
	Adder32 & $2.5\times10^{-4}$ & $15.8$ & $12.9$\\
	\hline
	Mult8 & $2.1\times10^{-3}$ & $13.8$ & $11.3$\\
	\hline
	Mult16 & $6.5\times10^{-4}$ & $15.6$ & $14.3$\\
	\hline
	Mult32 & $1.3\times10^{-5}$ & $14.7$ & $14.4$\\
	\hline
	Sobel & $2.0\times10^{-2}$ & $13.1$ & $12.6$\\
	\hline
	Fir8 & $3.0\times10^{-3}$ & $16.1$ & $14.5$\\
	\hline
	Blur & $2.3\times10^{-2}$ & $16.6$ & $16.5$\\
	\hline
	Mammogr & $1.3\times10^{-3}$ & $23.3$ & $22.5$\\
	\hline
	Vertebral & $9.2\times10^{-3}$ & $15.5$ & $17.2$\\
	\hline
	Balance & $9.5\times10^{-4}$ & $19.5$ & $20.8$\\
	\hline
    Redwine & $1.0\times10^{-2}$ & $13.9$ & $13.5$\\
	\hline
	Whitewine & $4.3\times10^{-3}$ & $15.1$ & $12.7$\\
	\hline
	Seeds & $1.8\times10^{-3}$ & $16.6$ & $19.0$\\
	\hline
	Cardio & $6.0\times10^{-4}$ & $11.5$ & $9.8$\\
	\hline
	\hline
	\textbf{Average} & $\bm{5.3\times10^{-3}}$ & $\bm{16.4}$ & $\bm{16.0}$\\
	\hline
\end{tabular}
}
\label{tab:ax_stats}
\end{table}
First, we thoroughly evaluate the delay and error characteristics of the approximate circuits generated by our framework to assess its effectiveness in eliminating timing violations caused by process variations, and thus, eradicating pessimistic timing guardbands.
As aforementioned, our heuristic optimization procedure produces a set of close-to-Pareto optimal approximate circuits, each featuring a unique trade-off between functional error due to logic approximation and CPD (see Section~\ref{sec:methodology}).
For each circuit, the extracted approximate netlists are evaluated by performing a $1000$-point Monte-Carlo analysis using our variability-aware standard cell libraries (for the 32-bit multiplier a $100$-point Monte-Carlo was used due to the elevated time of its evaluation cycle).
For each Monte-Carlo point we gather the CPD of the circuit via STA.
Overall, more than $3 \times 10^5$ design points are evaluated in our analysis.

Fig.~\ref{fig:pareto} presents the results of the Monte-Carlo analysis obtained from the final approximate solutions for all examined circuits.
For each examined circuit (i.e., subfigure in Fig.~\ref{fig:pareto}), the approximate designs generated by our framework are represented by a triangle (of any color).
For each approximate design, y-axis reports the purely functional error due to the functional approximation introduced by our framework.
Similarly, for each approximate design, x-axis reports its maximum CPD as obtained from the $1000$-point Monte-Carlo analysis.
Moreover, in each subfigure, the worst-case NMED of the baseline circuit (obtained from the $1000$-point Monte-Carlo, see Fig.~\ref{fig:wc_nmed}) is also included for comparative purposes (horizontal red dashed line). 
In addition, the nominal CPD of each baseline is also illustrated by a vertical red dotted line.
Overall, as shown in Fig.~\ref{fig:pareto}, our variability-aware designs offer significant advantages:
\begin{enumerate}
    \item
    They feature a considerable performance gain, as their Monte-Carlo-obtained \textit{worst-case} CPD is lower than the \textit{nominal} CPD of the baseline circuit.
    \item
    They \emph{eliminate} variability-induced quality degradation (output errors) as they can sustain the nominal frequency of the baseline without any timing violations.
    \item
    Under iso-performance conditions (i.e., when both baseline and approximate circuits are operated at the nominal frequency of the baseline), our approximate circuits feature \textit{improved accuracy} since their worst-case NMED value is lower that the worst-case NMED of the baseline.
    \item 
    Under iso-performance conditions, the error of our approximate circuits is only functional, whereas the baseline exhibits timing errors.
\end{enumerate}
An illustrative example of these advantages is presented in the subfigure of the Blur filter for a selected solution of the obtained Pareto front.
All the above demonstrate the effectiveness of our framework in accelerating selected, potentially critical paths, and thus, decreasing the circuit's CPD and its variation.
Each Pareto front comprises several approximate designs and covers a wide spectrum of possible trade-offs between CPD reduction w.r.t. the baseline circuit and injected functional error.
Hence, our framework offers the flexibility of selecting variability-aware approximate circuits with different combinations of latency and minimal functional error.

Targeting high accuracy, we select the approximate design with the minimum NMED (and worst-case CPD lower than the nominal delay of the baseline).
These designs are colored in orange in Fig.~\ref{fig:pareto} and evaluated thoroughly in Table~\ref{tab:ax_stats}.
Table~\ref{tab:ax_stats} reports the functional error due to approximation, the average CPD reduction and the reduction of the average standard deviation of the CPD of the selected approximate designs.
The reported reductions are calculated w.r.t. the corresponding values of the respective baseline circuit. 
Overall, we achieve a CPD reduction ranging from $11.5\%$ to $26.4\%$, for all examined circuits.
Furthermore, the standard deviation of the CPD has also been notably diminished, by $16.0\%$ on average for the examined circuits.
This reflects the importance of the three-fold objective minimization during our optimization procedure (see Section~\ref{sec:genetic}), where the reduction of the standard deviation of the delay distribution of potentially critical paths is directly targeted.
Hence, gates of wider delay distribution (i.e., increased standard deviation) may be consistently targeted for approximation by our genetic algorithm.
On the contrary, gates with low delay standard deviation might not be approximated, allowing, thus, their (unaffected) propagation paths to deliver proportionally reduced delay variability (i.e., standard deviation of the accumulated delay distribution).

\begin{figure}[t]
    \centering
    \resizebox{1\columnwidth}{!}{
        \includegraphics{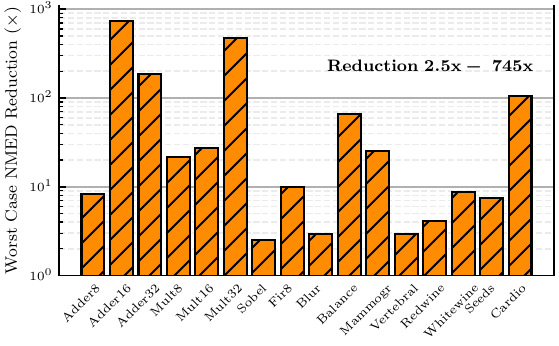}
    }
    \caption{Comparative evaluation of worst-case NMED of our approximate circuits against the baseline circuits. A $1000$-point Monte-Carlo analysis for all examined circuits is considered and the orange approximate designs of Fig.~\ref{fig:pareto} are used.}
    \label{fig:nmed_barplot}
\end{figure}

By inducing minimal functional error, our framework successfully suppresses unpredictable timing violations.
Fig.~\ref{fig:nmed_barplot} presents the worst-case error (NMED) reduction of our approximate designs of Table~\ref{tab:ax_stats} (and highlighted in orange in Fig.~\ref{fig:pareto}) w.r.t. the corresponding baseline circuits, when both are operated at the nominal delay of the baseline.
Note that for our circuits the worst-case NMED is identical to the approximation-induced NMED, since we eliminated the timing errors (i.e., our circuits feature purely functional errors).
Across all examined circuits, our generated designs exhibit significantly lower NMED, even for circuits whose worst-case NMED in Fig.~\ref{fig:wc_nmed} is particularly low (e.g., 32-bit multiplier).
In addition, we must note that the error of the baseline circuits originates from the timing violations and thus is unpredictable and cannot be bounded.
On the other hand, we induce a small functional error known at design time.
Overall, our NMED reduction ranges from $2.5\times$ to $745\times$.

\subsection{State of the Art Comparison}\label{sec:eval_sota}
\begin{figure}
    \centering
    \subfloat[]{
    \resizebox{1\columnwidth}{!}{
    \includegraphics{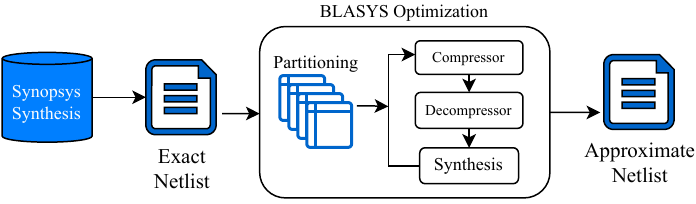}
    }
    \label{fig:sota_blasys}
    }
    \\
    \subfloat[]{
    \resizebox{1\columnwidth}{!}{
    \includegraphics{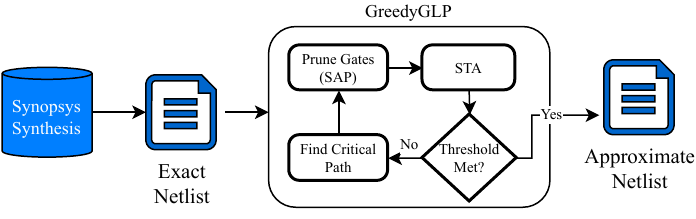}
    }
    \label{fig:sota_greedy}
    }
    \caption{Evaluation flow of the considered state-of-the-art techniques: (a) BLASYS and (b) GreedyGLP. The same baseline netlist (as obtained from Synopsys Design Compiler) is given as input to all frameworks.}
    \label{fig:sota_flow}
\end{figure}

In this section, we compare our framework against relevant state-of-the-art approximation techniques.
Specifically, we compare against the open-source BLASYS approximation framework~\cite{Hashemi:DAC:2018:blasys} which utilizes boolean matrix factorization to approximate the truth table of a given design.
BLASYS explores the design space of approximate, decomposed, and less complex subcircuits to obtain hardware efficiency with minimal accuracy loss due to approximations.
A simplified version of the evaluation flow for the BLASYS framework can be seen in Fig.~\ref{fig:sota_blasys}.

Furthermore, our framework is evaluated against a greedy GLP-based~\cite{Schlachter:IEEE-VLSI:2017:glp} delay optimization method for approximate circuits, namely GreedyGLP.
As aforementioned in Section~\ref{sec:related}, most approximate frameworks that target variability, direct the injected approximations to the critical path~\cite{Tsiokanos:DATE:2019:prec, Faryabi:EWDTS:2015, Zhang:GLSVLSI:2020, Gomez:TODAES:2018:dfs}.
However, remember that the CP is obtained via STA for a single degradation scenario.
GreedyGLP operates in a similar manner, by iteratively selecting and approximating the CP of a given gate-level netlist (with STA from Synopsys PrimeTime) until a given latency threshold is met.
The threshold enforces the CPD of the approximate circuit to be lower than the one of the respective baseline.
The criterion for selecting which gates of the CP will be approximated is obtained from GLP~\cite{Schlachter:IEEE-VLSI:2017:glp} (i.e., a significance-activity criterion is used), which is shown to retain low error magnitude and potentially, low error frequency.
Fig.~\ref{fig:sota_greedy} depicts the flow of GreedyGLP.

\textbf{To ensure a fair comparison}, all approaches (ours, BLASYS, and GreedyGLP) receive the same input, i.e., the Synopsys-optimized gate-level netlist of the baseline circuit (as illustrated in Fig.~\ref{fig:methodology} and \ref{fig:sota_flow}).
Both state-of-the-art methodologies output an approximate gate-level netlist, from their respective optimizations.
Similarly to our framework, the obtained netlists are synthesized with Design Compiler targeting zero-slack.
Then, we conduct a Monte-Carlo analysis using our $1000$ variability-aware libraries to evaluate the true characteristics of the approximate circuits (i.e., CPD and its standard deviation as well as their error) under variability degradations and thus enable fair comparison with our framework.
During the Monte-Carlo analysis we run both STA with PrimeTime and gate-level timing simulations (at the CPD of the baseline circuit) with Questasim.
STA is used to measure the CPD and its standard deviation, whereas the gate-level simulation provides the error of the approximate circuit (NMED).
For each related framework and each circuit, we repeat the above cycle for each of our $1000$ variability-aware standard cell libraries, resulting in an additional $32 \times 10^3$ delay-error evaluations.

\begin{figure}
    \centering
    \includegraphics{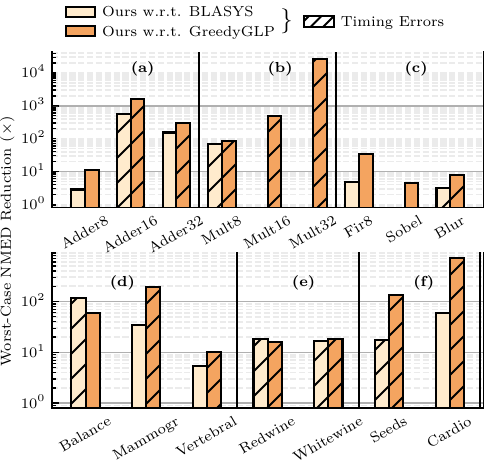}
    \caption{Worst-case NMED reduction of our framework w.r.t. the state of the art. The examined circuits are organized in (a) adders, (b) multipliers (c) image processing benchmarks, (d) MLPs, (e) SVMs and (f) DTs. Annotated bars indicate circuits that exhibited timing errors. BLASYS did not converge to an approximate netlist for the 16- and 32-bit multipliers and Sobel circuits.}
    \label{fig:sota_nmed}
\end{figure}

Fig.~\ref{fig:sota_nmed} compares the worst-case NMED attained by the approximate circuits generated by our framework against the one achieved by the approximate circuits generated by the the state-of-the-art BLASYS or GreedyGLP (as obtained from the respective Monte-Carlo simulations).
In Fig.~\ref{fig:sota_nmed}, designs which feature timing errors due to process variations are annotated with a diagonal pattern.
We should note that after $>48$h BLASYS did not converge to an approximate netlist for the 16- and 32-bit multipliers and Sobel circuit.
Overall, the approximate circuits generated by our framework feature significant NMED reduction over the state of the art.
Moreover, as shown in Fig.~\ref{fig:sota_nmed}, approximate circuits produced by BLASYS or GreedyGLP may suffer from variability-induced timing errors.
Such violations originate from the CP calculation of the relevant frameworks, which is obtained from a variability-unaware STA on the respective approximate circuit.
As demonstrated in Section~\ref{sec:motivation}, this approach cannot ensure that process-induced timing errors will be avoided.
Remember, our designs eliminate timing errors and the reported worst-case NMED is purely functional.
On average, our variability-aware circuits exhibit $79\times$ and $1936\times$ lower worst-case NMED, compared to BLASYS and GreedyGLP, respectively.

Note that, variability effects become more prominent for technology nodes of smaller transistor sizes~\cite{Huard:IRPS:2015:bti} (e.g. $7$\,nm) and/or different structures~\cite{FinFET_var}.
For such technologies, our framework will become more impactful.
With elevated gate variability, the phenomena described in the motivational examples in Section~\ref{sec:motivation} and Fig.~\ref{fig:motiv_example_simple}-\ref{fig:motiv_example} will be magnified.
Consequently, more paths would become potentially critical, and our probabilistic critical path estimation would be given a wider choice of potential paths to target for acceleration (i.e., introduce approximations within those paths).
Thus, further delay optimizations could be achieved.
Hence, although the presented experimental results rely on the $14$\,nm FinFET technology, even better results are expected for technology nodes of higher induced variability.

\subsection{Execution Time}\label{sec:eval_exec_time}
\begin{figure}
    \centering
    \resizebox{1\columnwidth}{!}{
    \includegraphics{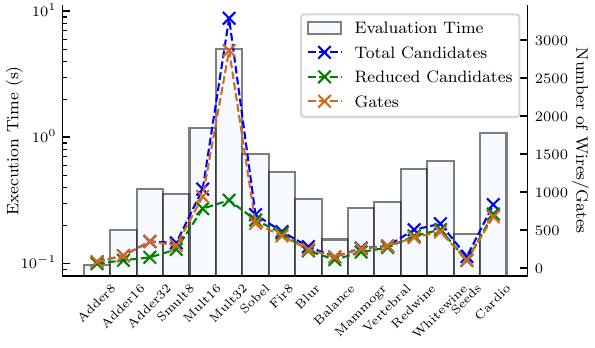}
    }
    \caption{Execution time of our framework for a single offspring evaluation (bars) for all examined circuits. The number of gates (orange dashed line), wires, i.e., original approximation candidates (blue dashed line) and reduced approximation candidates (green dashed line) are also included.}
    \label{fig:exec}
\end{figure}

Finally, we evaluate the execution time of our framework, and specifically, our evolutionary optimization procedure.
Our framework utilizes the NSGA-II algorithm, built upon high level estimators and thus, can be full parallelized.
As such, the time complexity of our framework is determined by user-defined parameters, such as the population size and number of evolution generations, the number of available threads of the underlying platform and the evaluation of a single approximate chromosome.
Hence, the execution time of a single offspring evaluation is selected for evaluation, in order to better characterize the time efficiency of our framework, since all other factors increase the scalability to more solutions.
Evaluating an approximate solution involves estimating its delay distribution by stochastically traversing the netlist-equivalent DAG, using the methodology described in Section~\ref{sec:delay}, as well as calculating its functional error due to approximation.
The histogram of Fig.~\ref{fig:exec} presents the execution time of a single offspring evaluation for all examined circuits.
The finalized measurements consist of the average of 100 trials and were conducted on a desktop computer featuring an AMD Ryzen 7 2700X 8-Core processor at 3.7 GHz and 32 GB of RAM.
Overall, the reported measurements remain low across all examined circuits.
The maximum execution time can be traced to our largest examined circuit, the 32-bit multiplier ($5.04s$), due to the strong correlation between the size of the circuit (i.e., the number of its comprising gates/wires) and the execution time of a single evaluation.
Even though a circuit's large size does not lead to prohibitively long evaluations, the enlarged design space caused be the increased amount of wires (i.e., approximation candidates) could potentially limit the exploration capabilities of the genetic algorithm.
Though, this inefficiency is significantly mitigated by the design space pruning scheme described in Section~\ref{sec:candidates}.
As demonstrated by the green line of Fig.~\ref{fig:exec}, the approximation candidates which determine the length of a single chromosome are kept at bay across all designs (e.g., for the 32-bit multiplier, only $27\%$ of the wires are retained as approximating candidates), thus preventing the design space from increasing proportionally with the netlist size.
Therefore, solutions of poor quality are not considered and a faster convergence of our optimization procedure due to the pruned design space is ensured.

\section{Conclusion}\label{sec:conclusion}
In this work, we propose an automated circuit-agnostic framework for generating variability-aware approximate circuits without timing guardbands.
Our framework explores the design space of possible approximate circuits with reduced delay variability and minimal error.
We accurately evaluate the effects of process variations on circuit delay with custom-built variability-aware libraries.
Targeting the elimination of timing guardbands due to variability, we demonstrate that by injecting negligible, purely functional approximation error at design time, we eliminate timing violations and increase robustness across the examined variability spectrum.

\begin{IEEEbiography}[{\includegraphics[width=1in,height=1.25in,clip,keepaspectratio]{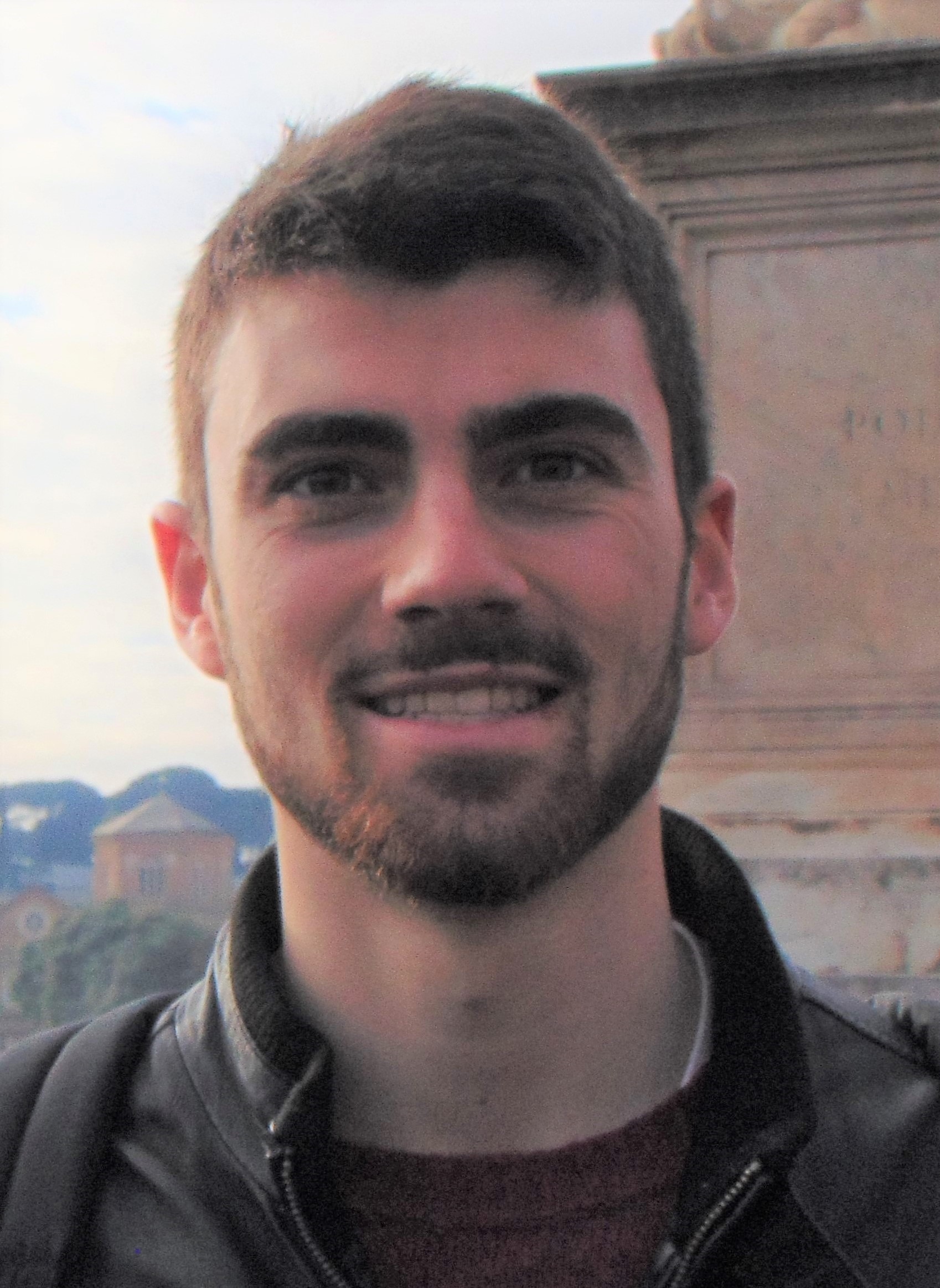}}]{Konstantinos Balaskas} received his Bachelor Degree in Physics and Master Degree in Electronic Physics from the Aristotle University of Thessaloniki in 2018 and 2020, respectively. Currently, he is a pursuing the PhD degree at the same institution. His research interests include approximate computing, machine learning and digital circuit design and optimization.
\end{IEEEbiography}

\begin{IEEEbiography}[{\includegraphics[width=1in,height=1.25in,clip,keepaspectratio]{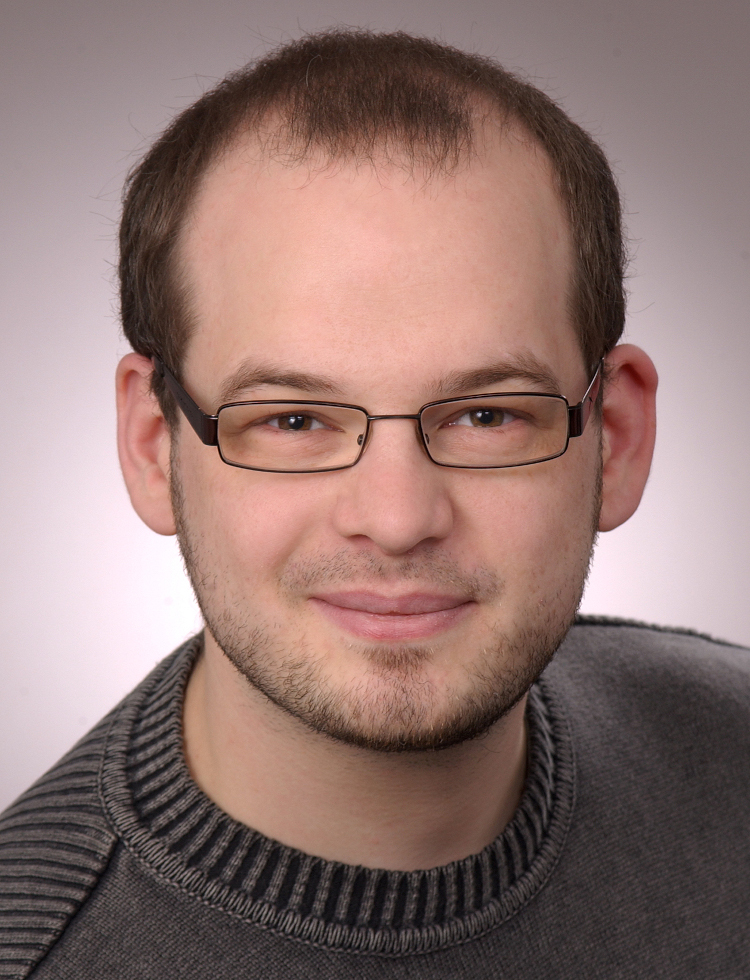}}] {Florian Klemme}(M'20) is a Doctoral Researcher at the Chair of Semiconductor Test and Reliability (STAR), University of Stuttgart.
He received the B.Sc. in System Integration from the University of Applied Sciences Bremerhaven, Germany, in 2014 and the M.Sc. in Computer Science from the Karlsruhe Institute of Technology, Germany, in 2018. He is currently working towards the Ph.D. degree at the Chair of Semiconductor Test and Reliability, University of Stuttgart. His research interests include cell library characterization and machine learning techniques in electronic design automation and computer-aided design.
He is a member of the IEEE. ORCID 0000-0002-0148-0523.
\end{IEEEbiography}

\begin{IEEEbiography}[{\includegraphics[width=1in,height=1.25in,clip,keepaspectratio]{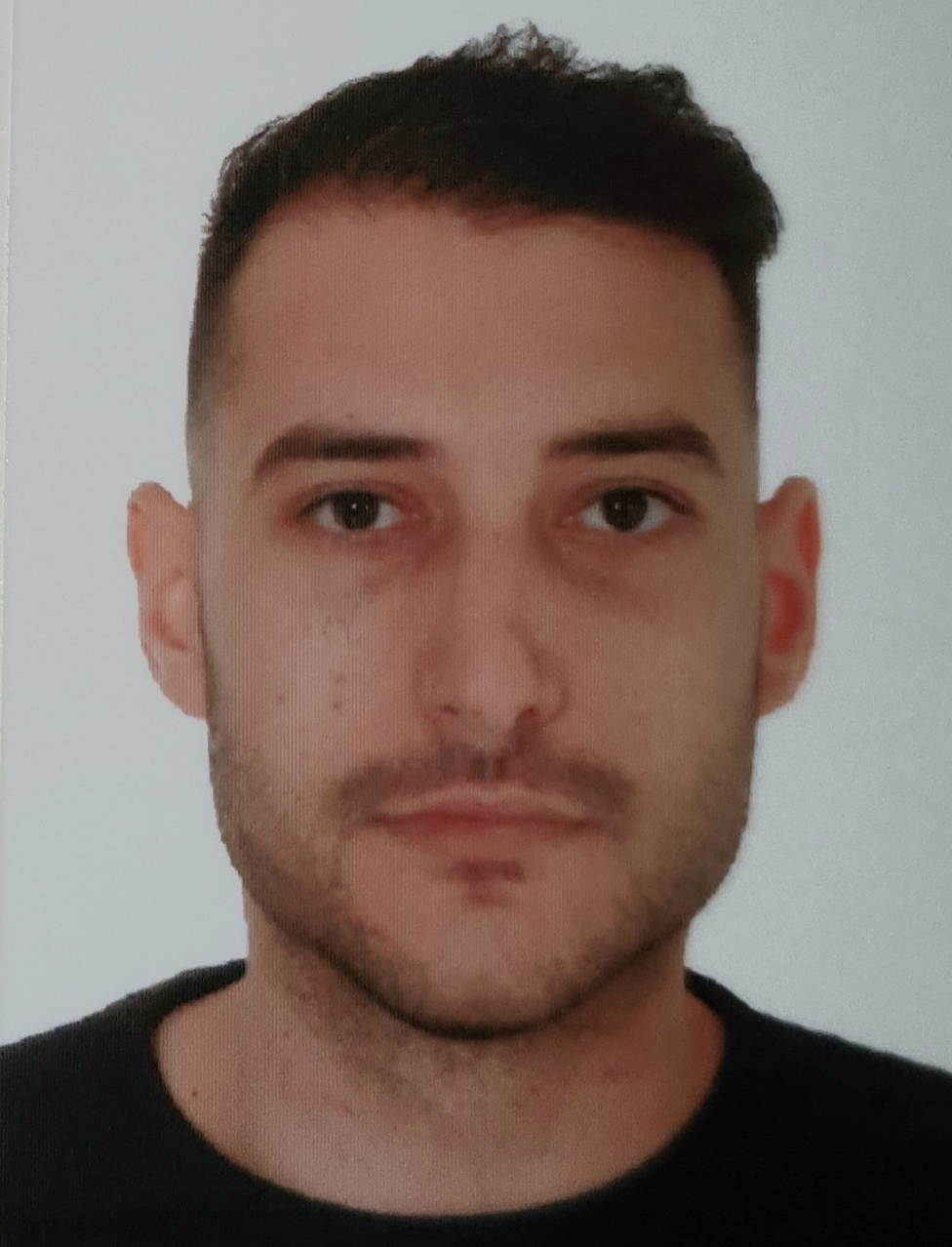}}] {Georgios Zervakis} is a Research Group Leader at the Chair for Embedded Systems (CES) at the Karlsruhe Institute of Technology (KIT), Germany.
He received the Diploma and the Ph.D. degree from the Department of Electrical and Computer Engineering (ECE), National Technical University of Athens (NTUA), Greece, in 2012 and 2018, respectively.
Before joining KIT, Georgios worked as a primary researcher in several EU-funded projects as member of the Institute of Communication and Computer Systems (ICCS), Athens, Greece.
His research interests include approximate computing, low power design, design automation, and integration of hardware acceleration in cloud.
\end{IEEEbiography}

\begin{IEEEbiography}[{\includegraphics[width=1in,height=1.25in,clip,keepaspectratio]{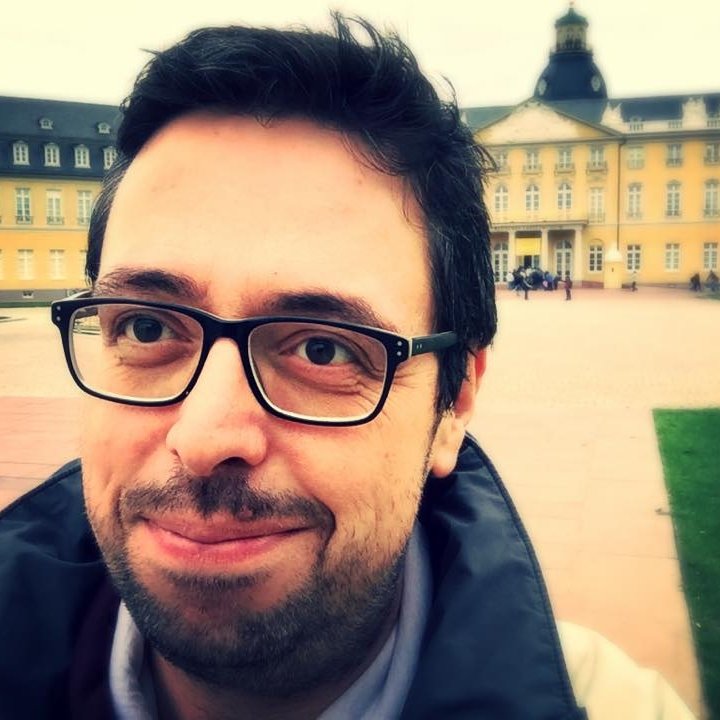}}]{Kostas Siozios} received his Diploma, Master and Ph.D. Degree in Electrical and Computer Engineering from the Democritus University of Thrace, Greece, in 2001, 2003 and 2009, respectively. Currently, he is Assistant Professor at Department of Physics, Aristotle University of Thessaloniki. His research interests include Low-Power Hardware Accelerators, Resource Allocation, Machine Learning, Decision-Making Algorithms, Cyber-Physical Systems (CPS) and IoT for Smart-Grid. He has published more than 130 papers in peer-reviewed journals and conferences. Also, he has contributed in 5 books of Kluwer and Springer. The last years he works as Project Coordinator, Technical Manager or Principal Investigator in numerous research projects funded from the European Commission (EC), European Space Agency (ESA), as well as the Greek Government and Industry.
\end{IEEEbiography}

\begin{IEEEbiography}[{\includegraphics[width=1in,height=1.25in,clip,keepaspectratio]{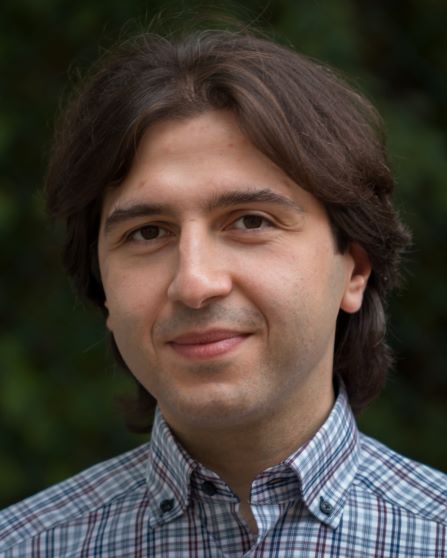}}] {Hussam Amrouch}(S'11-M'15) is a Junior Professor heading the Semiconductor Test and Reliability (STAR) chair within the Computer Science, Electrical Engineering Faculty at the University of Stuttgart as well as a Research Group Leader at the Karlsruhe Institute of Technology (KIT), Germany. He received his Ph.D. degree with highest distinction (Summa cum laude) from KIT in 2015. 
He servers currently as an Editor in the Nature Portfolio for the Nature Scientific Reports journal.
His main research interests are design for reliability and testing from device physics to systems, machine learning, security, approximate computing, and emerging technologies with a special focus on ferroelectric devices. He holds eight HiPEAC Paper Awards and three best paper nominations at top EDA conferences: DAC'16, DAC'17 and DATE'17 for his work on reliability. He also serves as Associate Editor at Integration, the VLSI Journal. He has served in the technical program committees of many major EDA conferences such as DAC, ASP-DAC, ICCAD, etc. and as a reviewer in many top journals like Nature Electronics, TED, TCAS-I, TVLSI, TCAD, TC, etc. He has 160+ publications in multidisciplinary research areas across the entire computing stack, starting from semiconductor physics to circuit design all the way up to computer-aided design and computer architecture. He is a member of the IEEE. ORCID 0000-0002-5649-3102.
\end{IEEEbiography} 

\begin{IEEEbiography}[{\includegraphics[width=1in,height=1.25in,clip,keepaspectratio]{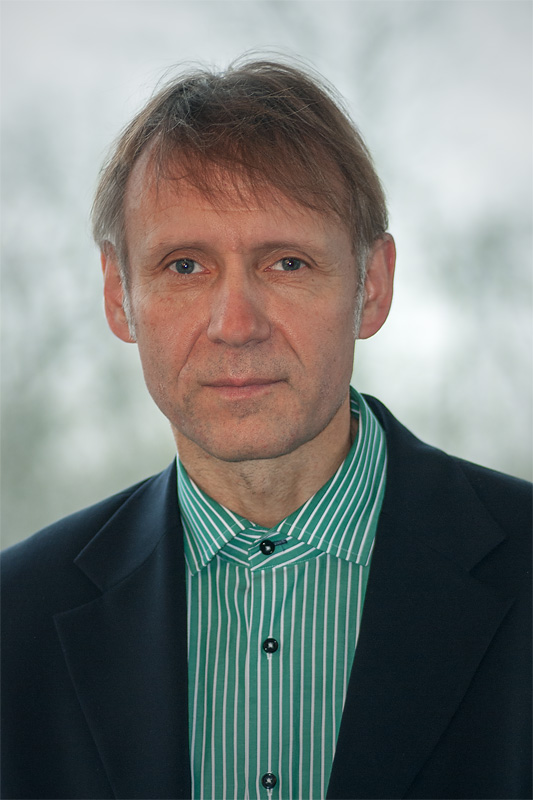}}]{J\"org Henkel} (M'95-SM'01-F'15)
is the Chair Professor for Embedded Systems at Karlsruhe Institute of Technology. Before that he was a research staff member at NEC Laboratories in Princeton, NJ. He received his diploma and Ph.D. (Summa cum laude) from the Technical University of Braunschweig. His research work is focused on co-design for embedded hardware/software systems with respect to power, thermal and reliability aspects. He has received six best paper awards throughout his career from, among others, ICCAD, ESWeek and DATE. For two consecutive terms he served as the Editor-in-Chief for the ACM Transactions on Embedded Computing Systems. He is currently the Editor-in-Chief of the IEEE Design\&Test Magazine and is/has been an Associate Editor for major ACM and IEEE Journals. He has led several conferences as a General Chair incl. ICCAD, ESWeek and serves as a Steering Committee chair/member for leading conferences and journals for embedded and cyber-physical systems. Prof. Henkel coordinates the DFG program SPP 1500 ``Dependable Embedded Systems'' and is a site coordinator of the DFG TR89 collaborative research center on ``Invasive Computing''. He is the chairman of the IEEE Computer Society, Germany Chapter, and a Fellow of the IEEE.
\end{IEEEbiography}


\begin{thebibliography}{10}
\providecommand{\url}[1]{#1}
\csname url@samestyle\endcsname
\providecommand{\newblock}{\relax}
\providecommand{\bibinfo}[2]{#2}
\providecommand{\BIBentrySTDinterwordspacing}{\spaceskip=0pt\relax}
\providecommand{\BIBentryALTinterwordstretchfactor}{4}
\providecommand{\BIBentryALTinterwordspacing}{\spaceskip=\fontdimen2\font plus
\BIBentryALTinterwordstretchfactor\fontdimen3\font minus
  \fontdimen4\font\relax}
\providecommand{\BIBforeignlanguage}[2]{{%
\expandafter\ifx\csname l@#1\endcsname\relax
\typeout{** WARNING: IEEEtran.bst: No hyphenation pattern has been}%
\typeout{** loaded for the language `#1'. Using the pattern for}%
\typeout{** the default language instead.}%
\else
\language=\csname l@#1\endcsname
\fi
#2}}
\providecommand{\BIBdecl}{\relax}
\BIBdecl

\bibitem{Bowman:IEEE-SSC:2010:45}
K.~A. Bowman \emph{et~al.}, ``A 45 nm resilient microprocessor core for dynamic
  variation tolerance,'' \emph{IEEE Journal of Solid-State Circuits}, vol.~46,
  no.~1, pp. 194--208, 2011.

\bibitem{Bhunia:VLSID:2007}
S.~Bhunia, S.~Mukhopadhyay, and K.~Roy, ``Process variations and
  process-tolerant design,'' in \emph{Int. Conf. VLSI}, 2007, pp. 699--704.

\bibitem{Alam:IRPS:2011:reliability}
M.~A. Alam, K.~Roy, and C.~Augustine, ``Reliability-and process-variation aware
  design of integrated circuits—a broader perspective,'' in
  \emph{International Reliability Physics Symposium}, 2011, pp. 4A--1.

\bibitem{Tsiokanos:DATE:2019:prec}
I.~Tsiokanos, L.~Mukhanov, and G.~Karakonstantis, ``Low-power variation-aware
  cores based on dynamic data-dependent bitwidth truncation,'' in \emph{2019
  Design, Automation \& Test in Europe Conference \& Exhibition (DATE)}.\hskip
  1em plus 0.5em minus 0.4em\relax IEEE, 2019, pp. 698--703.

\bibitem{Balaskas:TCASI:2021}
K.~Balaskas, G.~Zervakis, H.~Amrouch, J.~Henkel, and K.~Siozios, ``Automated
  design approximation to overcome circuit aging,'' \emph{IEEE Transactions on
  Circuits and Systems I: Regular Papers}, pp. 1--12, 2021.

\bibitem{Boroujerdian:ICCD:2018:temperature}
B.~Boroujerdian, H.~Amrouch, J.~Henkel, and A.~Gerstlauer, ``Trading off
  temperature guardbands via adaptive approximations,'' in \emph{Int. Conf. on
  Computer Design (ICCD)}.\hskip 1em plus 0.5em minus 0.4em\relax IEEE, 2018,
  pp. 202--209.

\bibitem{vSanten:IEEE-TCASI:2019:modeling}
V.~M. van Santen, H.~Amrouch, and J.~Henkel, ``Modeling and mitigating
  time-dependent variability from the physical level to the circuit level,''
  \emph{IEEE Trans. Circuits Syst. I, Reg. Papers}, vol.~66, pp. 2671--2684,
  2019.

\bibitem{Zervakis:IEEE-Access:2020}
G.~Zervakis, H.~Amrouch, and J.~Henkel, ``Design automation of approximate
  circuits with runtime reconfigurable accuracy,'' \emph{IEEE access}, vol.~8,
  pp. 53\,522--53\,538, 2020.

\bibitem{Faryabi:EWDTS:2015}
M.~Faryabi, H.~Dorosti, M.~Modarressi, and S.~M. Fakhraie, ``Process
  variation-aware approximation for efficient timing management of digital
  circuits,'' in \emph{IEEE East-West Design \& Test Symposium}, 2015, pp.
  1--4.

\bibitem{Wang:VLSI:2017}
Y.~Wang, J.~Deng, Y.~Fang, H.~Li, and X.~Li, ``Resilience-aware frequency
  tuning for neural-network-based approximate computing chips,'' \emph{IEEE
  Trans. Very Large Scale Integr. (VLSI) Syst.}, vol.~25, no.~10, pp.
  2736--2748, 2017.

\bibitem{Zhang:GLSVLSI:2020}
Z.~Zhang \emph{et~al.}, ``Reliability-enhanced circuit design flow based on
  approximate logic synthesis,'' in \emph{Great Lakes Symposium on VLSI
  (GLSVLSI)}, 2020, pp. 71--76.

\bibitem{Gomez:TODAES:2018:dfs}
A.~F. Gomez and V.~Champac, ``Selection of critical paths for reliable
  frequency scaling under bti-aging considering workload uncertainty and
  process variations effects,'' \emph{ACM Transactions on Design Automation of
  Electronic Systems (TODAES)}, vol.~23, no.~3, pp. 1--21, 2018.

\bibitem{Brendler:VLSI-SoC:2018}
L.~H. Brendler, A.~L. Zimpeck, C.~Meinhardt, and R.~Reis, ``Evaluating the
  impact of process variability and radiation effects on different transistor
  arrangements,'' in \emph{Int. Conf. Very Large Scale Integration (VLSI-SoC)},
  2018, pp. 71--76.

\bibitem{Gomez:ET:2019:gate_sizing1}
A.~Gomez and V.~Champac, ``An efficient metric-guided gate sizing methodology
  for guardband reduction under process variations and aging effects,''
  \emph{Journal of Electronic Testing}, vol.~35, no.~1, pp. 87--100, 2019.

\bibitem{Ebrahimipour:EmTopComp:2020:gate_sizing2}
S.~M. Ebrahimipour, B.~Ghavami, and M.~Raji, ``A statistical gate sizing method
  for timing yield and lifetime reliability optimization of integrated
  circuits,'' \emph{IEEE Trans. Emerg. Topics Comput.}, vol.~9, pp. 759--773,
  2020.

\bibitem{Raji:Access:2021:voltage_scaling}
M.~Raji, R.~Mahmoudi, B.~Ghavami, and S.~Keshavarzi, ``Lifetime reliability
  improvement of nano-scale digital circuits using dual threshold voltage
  assignment,'' \emph{IEEE Access}, vol.~9, pp. 114\,120--114\,134, 2021.

\bibitem{vSanten:IEEE-TCASI:2017:unified}
V.~M. Van~Santen, J.~Martin-Martinez, H.~Amrouch, M.~M. Nafria, and J.~Henkel,
  ``Reliability in super- and near-threshold computing: A unified model of rtn,
  bti, and pv,'' \emph{IEEE Transactions on Circuits and Systems I: Regular
  Papers}, vol.~65, no.~1, pp. 293--306, 2017.

\bibitem{Amrouch:DAC:2017:towards}
H.~Amrouch, B.~Khaleghi, A.~Gerstlauer, and J.~Henkel, ``Towards aging-induced
  approximations,'' in \emph{Design Automation Conference}, 2017.

\bibitem{Kim:IEEE-TCASI:2020:aging}
H.~Kim \emph{et~al.}, ``Aging compensation with dynamic computation
  approximation,'' \emph{IEEE Transactions on Circuits and Systems I: Regular
  Papers}, vol.~67, no.~4, pp. 1319--1332, 2020.

\bibitem{synopsys}
Synopsys. (2021) {Synopsys EDA Tools}.

\bibitem{intel_data}
S.~Natarajan \emph{et~al.}, ``A 14nm logic technology featuring 2 nd-generation
  finfet, air-gapped interconnects, self-aligned double patterning and a 0.0588
  {$\mu$}m 2 sram cell size,'' in \emph{2014 IEEE International Electron
  Devices Meeting}.\hskip 1em plus 0.5em minus 0.4em\relax IEEE, 2014, pp.
  3--7.

\bibitem{BSIM-CMG}
J.~P. {Duarte} \emph{et~al.}, ``Bsim-cmg: Standard finfet compact model for
  advanced circuit design,'' in \emph{Conference 2015 - 41st European
  Solid-State Circuits Conference (ESSCIRC)}, 2015, pp. 196--201.

\bibitem{FinFET_var}
H.~Amrouch \emph{et~al.}, ``Impact of variability on processor performance in
  negative capacitance finfet technology,'' \emph{IEEE Transactions on Circuits
  and Systems I: Regular Papers}, vol.~67, no.~9, pp. 3127--3137, 2020.

\bibitem{nangate15}
M.~Martins \emph{et~al.}, ``Open cell library in 15nm freepdk technology,'' in
  \emph{Int. Symp. on Physical Design (ISPD)}, 2015, p. 171–178.

\bibitem{Silvaco}
Silvaco. {Silvaco Open Source Library}.

\bibitem{Siliconsmart}
Synopsys. {SiliconSmart}.

\bibitem{Schlachter:IEEE-VLSI:2017:glp}
J.~Schlachter, V.~Camus, K.~V. Palem, and C.~Enz, ``Design and applications of
  approximate circuits by gate-level pruning,'' \emph{IEEE Trans. Very Large
  Scale Integr. (VLSI) Syst.}, vol.~25, pp. 1694--1702, 2017.

\bibitem{Jain:DATE:2016:wbw}
S.~Jain, S.~Venkataramani, and A.~Raghunathan, ``Approximation through logic
  isolation for the design of quality configurable circuits,'' in \emph{2016
  Design, Automation \& Test in Europe (DATE)}.\hskip 1em plus 0.5em minus
  0.4em\relax IEEE, pp. 612--617.

\bibitem{Momeni:TC:2014:NMED}
A.~Momeni, J.~Han, P.~Montuschi, and F.~Lombardi, ``Design and analysis of
  approximate compressors for multiplication,'' \emph{IEEE Transactions on
  Computers}, vol.~64, no.~4, pp. 984--994, 2014.

\bibitem{Olya:Inderscience:2014:dijkstra}
M.~H. Olya, ``Applying dijkstra’s algorithm for general shortest path problem
  with normal probability distribution arc length,'' \emph{International
  Journal of Operational Research}, vol.~21, no.~2, pp. 143--154, 2014.

\bibitem{Deb:IEEE:2002:NSGAII}
K.~Deb, A.~Pratap, S.~Agarwal, and T.~Meyarivan, ``A fast and elitist
  multiobjective genetic algorithm: Nsga-ii,'' \emph{IEEE Transactions on
  Evolutionary Computation}, vol.~6, no.~2, pp. 182--197, 2002.

\bibitem{Zervakis:TVLSI:2019:vader}
G.~Zervakis, K.~Koliogeorgi, D.~Anagnostos, N.~Zompakis, and K.~Siozios,
  ``Vader: Voltage-driven netlist pruning for cross-layer approximate
  arithmetic circuits,'' \emph{IEEE Trans. on Very Large Scale Integration
  Systems}, vol.~27, no.~6, pp. 1460--1464, 2019.

\bibitem{Paim:TCASI:2021}
G.~{Paim} \emph{et~al.}, ``{On the Resiliency of NCFET Circuits Against Voltage
  Over-Scaling},'' \emph{IEEE Transactions on Circuits and Systems I: Regular
  Papers}, vol.~68, no.~4, pp. 1481--1492, 2021.

\bibitem{Lee:Springer2019:ahls}
S.~Lee and A.~Gerstlauer, \emph{Approximate High-Level Synthesis of Custom
  Hardware}.\hskip 1em plus 0.5em minus 0.4em\relax Springer International
  Publishing, 2019, pp. 205--223.

\bibitem{Dua:2019:datasets}
D.~Dua and C.~Graff, ``{UCI} machine learning repository,'' 2017.

\bibitem{Hashemi:DAC:2018:blasys}
S.~Hashemi, H.~Tann, and S.~Reda, ``Blasys: Approximate logic synthesis using
  boolean matrix factorization,'' in \emph{2018 55th ACM/ESDA/IEEE Design
  Automation Conference (DAC)}, 2018, pp. 1--6.

\bibitem{Huard:IRPS:2015:bti}
V.~Huard, D.~Angot, and F.~Cacho, ``From bti variability to product failure
  rate: A technology scaling perspective,'' in \emph{2015 IEEE International
  Reliability Physics Symposium}.\hskip 1em plus 0.5em minus 0.4em\relax IEEE,
  2015, pp. 6B--3.

\end{thebibliography}
\end{document}